# A Bio-Medical Snake Optimizer System Driven by Logarithmic Surviving Global Search for Optimizing Feature Selection and its application for Disorder Recognition

Ruba Abu Khurma,[a,b] Esraa Alhenawi,[c] Malik Braik,[d] Fatma A. Hashim,[e] Amit Chhabra,[f] Pedro A. Castillo [g]

[a] *MEU Research Unit, Faculty of Information Technology, Middle East University, Amman, 11831, Jordan. Faculty of Information Technology,*
[b] *Computer Science Department, Al- Ahliyya Amman University, Amman, Jordan, E-mail: r.khurma@ammanu.edu.jo*
[c] *Software Engineering Department, Al- Ahliyya Amman University, Amman, Jordan, E-mail: e.alhenawi@ammanu.edu.jo*
[d] *Computer Science Department, Prince Abdullah bin Ghazi Faculty of Information and Communication Technology, Al-Balqa Applied University, Salt, Jordan.*
[e] *Faculty of Engineering, Helwan University, Egypt. E-mail: fatma_hashim@h- eng.helwan.edu.eg*
[f] *Department of Computer Engineering and Technology, Guru Nanak Dev University, Amritsar, 143005, India, amit.cse@gndu.ac.in*
[g] *Department of Computer Engineering, Automatics and Robotics, University of Granada, Granada, Spain, pacv@ugr.es*

**Abstract**

It is of paramount importance to enhance medical practices, given how important it is to protect human life. Medical therapy can be accelerated by automating patient prediction using machine learning techniques. To double the efficiency of classifiers, several preprocessing strategies must be adopted for their crucial duty in this field. Feature selection (FS) is one tool that has been used frequently to modify data and enhance classification outcomes by lowering the dimensionality of datasets. Excluded features are those that have a poor correlation coefficient with the label class, that is, they have no meaningful correlation with classification and do not indicate where the instance belongs. Along with the recurring features, which show a strong association with the remainder of the features. Contrarily, the model being produced during training is harmed, and the classifier is misled by their presence. This causes overfitting and increases algorithm complexity and processing time. The pattern is made clearer by FS, which also creates a broader classification model with a lower chance of overfitting in an acceptable amount of time and algorithmic complexity. To optimize the FS process, building wrappers must employ metaheuristic algorithms (MAs) as search algorithms. The best solution, which reflects the best subset of features within a particular medical dataset that aids in patient diagnosis, is sought in this study using the Snake Optimizer (SO). The swarm-based approaches that SO is founded on have left it with several general flaws, like local minimum trapping, early convergence, uneven exploration and exploitation, and early convergence. By employing the cosine function to calculate the separation between the present solution and the ideal solution, the logarithm operator was paired with SO to better the exploitation process and get over these restrictions. In order to get the best overall answer, this forces the solutions to spiral downward. Additionally, SO is employed to put the evolutionary algorithms' preservation of the best premise into practice. This is accomplished by utilizing three alternative selection systems tournament, proportional, and linear to improve the exploration phase. These are used in exploration to allow solutions to be found more thoroughly and in relation to a chosen solution than at random. TLSO, PLSO, and LLSO stand for Tournament Logarithmic Snake Optimizer, Proportional Logarithmic Snake Optimizer, and Linear Order Logarithmic Snake Optimizer, respectively. A number of 22 reference medical datasets were used in experiments. The findings indicate that, among 86 % of the datasets, TLSO attained the best accuracy, and among 82 % of the datasets, the best feature reduction. In terms of the standard deviation, the TLSO also attained noteworthy reliability and stability. On the basis of running duration, it is, nonetheless, quite effective.

*Keywords*: Snake optimizer, Evolutionary algorithms, Selection systems, Survival of the fittest principle, Logarithmic function

# 1. Introduction

The right diagnosis of a health condition is a critical step in the medical process since it determines the patient's course of therapy and makes it possible to administer that treatment. Microarrays, medical imaging, and biomedical sig- nal processing datasets are the three main categories of medical data. Various kinds of data sets are unstructured, but microarray data is a prime instance of structured data collection. Medical datasets pack a variety of noise features, which comprise recurring and ineffective features J. Li et al. (2017). Such features have a detrimental influence on the categorization process. Feature selection (FS), which discards noise features from high-dimensional datasets and identifies the most extremely helpful feature set for use in the course of treatment and diagnosis.

Based on the presence of label classes, the literature provides a variety of FS techniques that may be broken down into three different categories: supervised, semi-supervised, and unsupervised techniques Diao and Shen (2015). The presence of a specific label for the intended class in the dataset identifies the supervised category, while the semi-supervised and unsupervised categories are established by having all of the examples or certain examples have the label. On the other hand, the five basic categories into which FS techniques can be grouped depending on the method of selection are filters, wrappers, embedding, ensemble, and combinations of these. Before selecting a subset of the features with a high rank value, each feature is given a rank using the statistical method known as filtering. Filters are evaluated using some statistical tests, as opposed to wrappers, where evaluation is based on classifier performance. Wrappers select the subset of features that delivers the best fitness value out of all the subsets of features generated by the search algorithm used in the feature deployment Gao, Wang, and Pedrycz (2020); M. Li and Wang (2022); Wang, Gao, and Pedrycz (2022).

Since the selection process is incorporated into the classifier, the integrated FS technique automatically selects the set of features that yield the best results Ang, Mirzal, Haron, and Hamed (2015). When discussing the usage of various FS techniques on the same dataset, respectively, both the terms "homogeneous" and "heterogeneous" are used. This is particularly true for ensemble FS techniques Rouhi and Nezamabadi-Pour (2020). The hybrid feature selection techniques are the most prevalent. To be able to take full advantage of the benefits of each technique inevitably as well as enhance classification performance, this assortment of techniques focuses on bringing together two or more FS techniques from several selection techniques, particularly filters and wrappers Manikandan and Abirami (2018).

According to the data perspective, FS methods can also be divided into categories such as FS methods for regular data, feature structure-based methods, heterogeneous data methods, and streaming data methods Khurma, Awadallah, and Aljarah (2021). For the regular data category, all features have been treated as independent features, and the significance of each feature has been determined using either the similarity of features, some heuristic criteria of filters, or depending on a specific statistical measure Khurma, Castillo, Sharieh, and Aljarah (2020). Structure-based FS techniques are FS techniques that improve classification results by taking into account the feature structure throughout FS. Nevertheless, FS techniques for managing data from several sources, or viewpoints, might be identified under the heterogeneous data methods category. The category of streaming data methods covers FS techniques that are capable of being used for streaming Khurma et al. (2023).

The main research problem in this study is to minimize the number of selected features and at the same time achieve high accuracy results for classification of diseases. Based on this stated research problem, we can put a number of research questions:

1. How we can minimize the number of selected features in a medical dataset and simultaneously achieve high classification results?
2. What is the suitable search algorithm to be adopted for handling FS process?
3. What are the enhancement operators that may be needed to be hybridized with the search algorithm to enhance its performance?
4. What are the evaluation measures that need to be applied to test the capability of the generated algorithm?

Search algorithms are highly related to the FS process Khurma, Aljarah, and Sharieh (2021), as in general, any FS method tries to search for the most informative set of features by deploying specific search algorithms that are usually represented by a metaheuristic search algorithm (MA) Khurma, Al-jarah, and Sharieh (2020b). MAs are higher-level algorithms that are used for generating optimal or near-optimal solutions for optimization problems. MAs involve two categories: evolutionary and swarm intelligence algorithms Khurma, Aljarah, and Sharieh (2020a).

On the whole, each FS perspective strives to discover the most meaningful combination of features by adopting specific search algorithms, which typically correspond with an MA. Therefore, search algorithms have a profound connection to the FS process Abualigah and Dulaimi (2021); Ewees et al. (2021); Lappas and Yannacopoulos (2021); Maleki, Zeinali, and Niaki (2021). MAs are higher level algorithms that result in optimal or near-optimal solutions to optimization problems. Swarm intelligence and evolutionary algorithms are two distinct categories of MAs. Motivated by the natural progression of organisms, evolutionary algorithms attempt to identify the optimum solution. This is a solution that offers a superior fitness value among the population of alternatives to consider to a given optimization problem.

The snake optimizer (SO), is a freshly developed swarm-based technique for taking care of various optimization concerns. We chose SO owing to its high stability, uncomplicated structure, outstanding convergence behavior, and parameter-free method. Even so, plenty of swarm-based algorithms additionally come with deficiencies, notably the capture of local minima, inequitable exploration, exploitation stages, and exaggerated convergence. The SO algo- rithm's amazing properties and the optimization results obtained by SO in the literature encouraged us to propose it for conducting optimization processes within the wrapper FS framework. Furthermore, the significant aspects of the classification result in the medical diagnosis sector necessitating some opera- tors to deal with the SO's shortcomings and improve its classification power, such as distinguishing between malevolent and benign tumors with high ac- curacy. To remedy these numerous deficiencies, several kinds of adjustment methods have been incorporated into the SO framework in this study. Especially since SO is utilized as a search algorithm in the wrapper framework for enhancing the procedure of diagnosing diseases. The study's fundamental accomplishments are:

- Generates a binary version of the SO algorithm (referred to as BSO) to take on the FS process utilizing the sigmoid function.
- Improve the exploitation process of the SO algorithm by using the logarithmic cosine function that is borrowed from the Moth Flame Optimizer Abu Khurmaa, Aljarah, and Sharieh (2021) which is called LBSO.
- Improve the exploration process of the SO by adopting three evolutionary selection operators and applying the survival of the fittest principle. These new versions are called Tournament Logarithmic Snake Optimizer (TLBSO), Proportional Logarithmic Snake Optimizer (PLBSO), and Linear Rank Logarithmic Snake Optimizer (LLBSO).
- Applying the improved BSO in the medical application for enhancing the disease diagnosis process by using 22 medical datasets.

The remaining portions of the paper have been laid out in the following fashion: Sec. 2 reviews a handful of the most recently published studies that used MA-FS techniques for medical diagnosis. The technical specifications of the snake optimizer are given in Sec. 3. The background details for the binary SO and SO for FS are presented in Sec. 4. The recommended approach is described in Sec. 5. Offered. In the Section titled Sec. 6, the experiments, outcomes, thorough examination, and argument are presented. Sec. 7, the conclusion wraps up the paper by outlining the paper's significant themes while offering an overview of a few upcoming endeavors.

## 2. Literature review

Numerous approaches based on MAs wrappers that have been implemented for healthcare applications are widely available in the literature. Readers may find out additional information regarding MA-FS in a survey Abu Khurma et al. (2022). Yet, the No-Free-Lunch theorem Adam, Alexandropoulos, Parda- los, and Vrahatis (2019) prompted academics to carry on their investigations and design additional MA algorithms for dealing with other optimization chal- lenges. No MA can be proclaimed the most effective methodology for all opti- mization challenges according to the NFL Awadallah, Al-Betar, et al. (2022); Awadallah, Hammouri, Al-Betar, Braik, and Abd Elaziz (2022).

Ragunthar Ragunthar and Selvakumar (2019) combined stochastic diffusion search and artificial bee colony (ABC) methods to create an FS in bone marrow plasma cell gene expression data. They applied a support vector ma- chine (SVM) as a classifier to two datasets. The outcomes indicated that the proposed technique performed better than both the ABC and the original stochastic diffusion search alone. Jain created a wrapper approach utilizing a binary version of the particle swarm optimization (PSO) I. Jain, Jain, and Jain (2018). Six cancer datasets and three classifiers were employed. The proposed strategy escaped local minimum stagnation, which boosted classification accuracy. Punitha et al. Yan and Lu (2019) deployed a genetic algorithm (GA) as the search algorithm in the hybrid wrapper they proposed for cancer diagnostic applications. The proposed method outperformed some FS methods for cancer prediction. A genetic algorithm (GA) was used as the search algorithm in the hybrid wrapper that Punitha et al. presented for cancer diagnostic applications. For predicting cancer, the suggested technique performed better than some FS methods. Sathiyabhama Sathiyabhama et al. (2021) and colleagues conducted a further study for the prediction of breast cancer. Using a computerized diagnosis system called a grey wolf optimization (GWO) based on rough set (RS) theory. The performance of the GWORS system was compared to a few hybrid systems that combine rough sets with either GA or PSO. The suggested GWORS demonstrated improved accuracy and F-measure performance.

A hybrid FS approach based on community detection and GA was devel- oped by Rostami et al. and is known as CDGAFS. The feature similarities are computed in the first of three processes that make up this approach. The following steps involve grouping the characteristics using the community dis- covery algorithm and FS using GA. To compare CDGAFS to three other approaches-based on PSO (PSO-FS), ant colony optimization (ACO-FS), and artificial bee colony algorithms (ABC-FS), they employed six medical datasets with SVM, KNN, and AdaBoost classifiers. The suggested CDGAFS fared better than these three techniques in terms of accuracy, with an average improvement of 0.52 percent, 1.20 percent, and 1.57 percent, respectively Ros-tami, Berahmand, and Forouzandeh (2021). Jain et al. S. Jain and Dharavath (2021) deployed the modified binary version of an SSA for FS in agricultural disease detection by using a quadratic transfer function to build a binary ver- sion of the salp swarm optimization algorithm (SSA). Using eleven datasets, they evaluated the proposed algorithm's performance in comparison to five of the most recent FS algorithms. The outcomes showed that the suggested algo- rithm performed better than any other algorithm tested. To produce a binary MFO from the canonical one for medical datasets of various sizes, Nadimi- Shahraki et al. devised an FS approach based on three forms of moth-flame optimization (MFO) with different transfer functions, including S-shaped, V- shaped, and U-shaped transfer functions Nadimi-Shahraki, Banaie-Dezfouli, Zamani, Taghian, and Mirjalili (2021).

Some of the most recent research that published an intelligent water drop (IWD) algorithm in the FS process for breast cancer prediction and microarray data processing for cancer prediction are the studies Kalita, Singh, and Kumar (2022a), Alhenawi, Al-Sayyed, Hudaib, and Mirjalili (2023), and Kalita, Singh, and Kumar (2022b). The original IWD algorithm was used by Kalita et al. to remove extraneous features and retain only the relative features. To compare the effectiveness of the suggested strategy to other MA-FS methods, they used SVM and the Breast Image Analysis Association (MIAS) database for evaluation. With an accuracy of 97.9%, the IWD fared better than previous approaches described in the literature Kalita et al. (2022a). The authors at reference Kalita et al. (2022b) introduced a threshold-based two way IWD FS algorithm for early stage breast cancer prediction by selecting lower and higher limits in the same year. As the best subset was utilized to train the SVM classifier, this technique allows for the production of many best feature subsets as opposed to a single best subset. With an accuracy value of 99 percent, a specificity of 96.2 percent, and an F1 score of 98.4 percent, the findings demonstrated the efficiency of the suggested strategy on some of the most popular algorithms.

GA was one of the evolutionary computation methods most frequently used for FS in an illness diagnosis. The authors assessed various FS algorithms and classifier combinations, using mutual information gain, additional trees, and GA as FS algorithms, for the early identification of Parkinson's disease Lamba, Gulati, Alharbi, and Jain (2022). As a classifier, naive Bayes, k-nearest-neighbors, and random forests were employed. They handled the un- balancing issue in this dataset by using the speech dataset for evaluation and the SMOTE oversampling method. The outcomes showed that a GA and random forest combination had the best accuracy performance, coming in at 95.58%. Alweshah et al. Lamba et al. (2022) deployed two wrapper methods, the first of which used the coronavirus herd immunity metaheuristic algorithm (CHIO), and the second of which used a combination of CHIO and greedy crossover (GC) to enhance the global search capability of the CHIO algorithm. For evaluation, they employed 23 actual medical datasets along with COVID-19. The outcomes showed that, depending on accuracy, selection size, and F-measure, CHIO-GC outperformed CHIO. Additionally, CHIO-GC outperformed six FS approaches from the literature. GA was used as a component of a hybrid feature selection technique they created for cancer detection. Using two classifiers and thirteen microarray datasets, they compared the proposed method to other FS methods from the literature Deng, Li, Deng, and Wang (2022). Based on the accuracy, F-score, precision, and recall measures, the results showed that the suggested method outperformed all other methods that were tested. The WOA algorithm was converted to binary and enhanced by the authors Nadimi-Shahraki, Zamani, and Mirjalili (2022). When utilized to treat coronavirus illness, the suggested techniques outperformed several WOA iterations that were employed to address global optimization issues.

Concisely, there is still room for new optimizers and augmentation techniques. As a result, the FS process may be supported, and near optimal solutions that maximize performance can be produced. This inspired us to suggest SO with survival exploration operators and a logarithmic spiral method. Use a sufficient amount of medical datasets to apply this methodology to identify the disease. To the best of our knowledge, this optimizer has never been utilized in medical applications and hasn't been optimized using these operators.

## 3. Snake Optimizer (SO)
### 3.1. Inspiration from nature

In Hashim and Hussien (2022), Hashem used the natural behavior of snakes to construct a mathematical optimization model. The life cycle of snakes is fascinating because of how they affect the surrounding natural elements. Fig. 1 shows different stages of snake life. Fig. 1.1 is the snake. Fig. 1.2 depicts the close fights between snakes, Fig. 1.3 shows the snake mating, and Fig. 1.4 shows the snakes laying egg stages. The snakes will devour the food if it is available and the air is heated. However, the snakes mate when food is present and the air is frigid.

When there is food and a cold temperature in the air, snakes mate. In addition, they look for food if these natural circumstances are not present. The SO model is based on the same two primary stages as the other swarm- based algorithms: global search (exploration) and local search (exploitation). When food is scarce and the weather is chilly, snakes in the exploration phase scatter throughout the foraging area in search of food. The exploitation stage, on the other hand, is broken down into several smaller

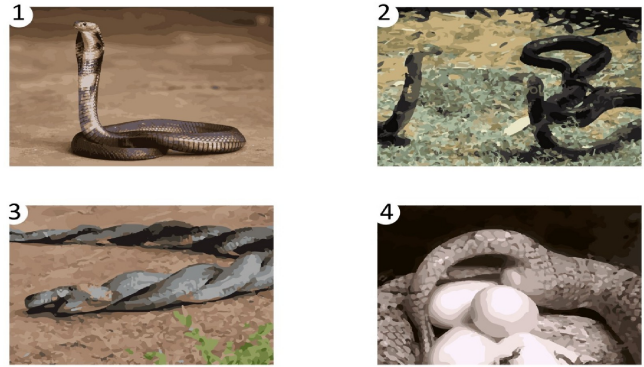

Fig. 1: Stages in the snakes' life

### 3.2. Mathematical Model

This section explains the mathematical model of an SO algorithm Hashim and Hussien (2022). The following steps illustrate the SO model:

– Initialization of solutions: In the search space, SO establishes a collection of random solutions utilizing Eq. (1). These solutions form the population of snakes that the SO will optimize in the next steps.

$$S_i = S_{min} + R \times (S_{max} - S_{min}) \quad (1)$$

where Si represents the location of the ith snake in the population within the search area. A random number in[0, 1] is called R. The optimization problem's lowest and highest possible values are Smax and Smin, respectively.

– Snake population division: Using Eq. (2) and Eq. (3), the population is split evenly into two sections (50 percent male and 50 percent female).

$$N_m = N/2 \quad (2)$$
$$N_f \approx N - N_m \quad (3)$$

where N is the total population size of all snakes. The total number of male snakes is $N_m$. The total number of female snakes is $N_f$.

– Evaluation of snakes: get the best snake from the male and female groups (Sbestm and Sbestf), and then locate the food ($L_{food}$). The definitions of two additional terms are temperature (T) and food quantity (Q), which are expressed respectively as Eq. (4) Temperature and Eq. (5) Quantity.

$$T = e^{\frac{-c}{C}} \quad (4)$$

where c is the current iteration and C is the maximum number of iterations.

$$Q = k_1 \times e^{\left(\frac{c-C}{c}\right)} \qquad (5)$$

where $K_1$ is a constant value equal to 0.5.

- Investigating the search area (no food is located): This is predicated on the use of a chosen threshold value. If Q 0.25, the snakes update their positions in relation to a given random position inside the search space in order to search globally. the model for this Eq. (6) - Eq. (9)

$$Sm_i(c+1) = Sm_R(c) \pm K_2 \times AB_m \times ((S_{max} - S_{min}) \times R + S_{min}) \qquad (6)$$

where R is a random number in[0, 1], Smi is the location of the ith male snake, and $Sm_R$ is the location of a random male snake. The male snake's capacity to locate food is represented by $AB_m$ and can be discovered using Eq. (7):

$$AB_m = e^{(-Fm\ R)/(\ Fm\ i)} \qquad (7)$$

where $K_2$ is a constant equal to 0.05 and $Fm_R$ is the fitness of the $Sm_R$ snake and Fmi is the fitness of the ith snake in the male group.

$$Sf_i(c+1) = Sf_R(c) \pm K_2 \times AB_f \times ((S_{max} - S_{min}) \times R + S_{min}) \qquad (8)$$

where R is a random number in the range of [0, 1], Sfi is the position of the ith female snake, SfR is the position of a random female snake, and ABf is the ability of the female snake to locate food. Eq. (9):

$$AB_f = e^{(-Ff\ R)/(\ Ff\ i)} \qquad (9)$$

where $Ff_R$ is the fitness of the snake in the male group $Sf_R$, Ffi is the fitness of the snake in the ith snake in the male group, and $K_2$ is a constant equal to 0.05.
- Exploitation the search area (Food is discovered) The temperature is tested if the amount of food is larger than a predetermined threshold Q > 0.25. The solutions will only travel to the food if T > 0.6 (hot).

$$S_{(i,j)}(c+1) = L_{food} \pm K_3 \times T \times R \times (L_{food} - S_{(i,j)}(c)) \qquad (10)$$

where S(i, j) represents a snake's (male or female) position, Lfood represents the best snakes, and K3 represents a constant value equal to 2.

If T > 0.6 (cold), The snake will be in the fight mode or mating mode Fight Mode.
– Fight mode

$$Sm_i(c+1) = Sm_i(c) \pm K_3 \times FAM \times R \times (Sf_{best} - Sm_i(c)) \qquad (11)$$

Where $Sm_i$ is the ith male position, Sfbest is the position of the best snake in the female group, and F AM is the fighting ability of the male snake.

$$Sf_i(c+1) = Sf_i(c) \pm K_3 \times FAF \times R \times (Sm_{best} - Sf_i(c)) \qquad (12)$$

where Sfi is the ith female position, Smbest is the position of the best snake in the male group, and FAF is the fighting ability of the female snake. F AM and F AF can be found from the following equations:

$$FAM = e^{(-Ffbest)/(Fi)} \quad (13)$$

$$FAF = e^{(-Fmbest)/(Fi)} \quad (14)$$

where Fmalebest signifies the best snake fitness for the male group, Fi denotes snake fitness, and Ffbest denotes the best snake fitness for the female group.

– Mating mode

$$Sm_i(c+1) = Sm_i(c) \pm K_3 \times MA_m \times R \times (Q \times Sf_i(c) - Sm_i(c)) \quad (15)$$

$$Sf_i(c+1) = Sf_i(c) \pm K_3 \times MA_{fm} \times R \times (Q \times Sm_i(c) - Sf_i(c)) \quad (16)$$

where Sfi and Smi are the places of the ith snake in the female and male groups, respectively, and MAm and MAf are the capacities of males and females for mating, respectively, and they may be derived as follows or found *as follows:*

$$MAm = e^{(-Ffi)/(Fmi)} \quad (17)$$

$$MAf = e^{(-Fmi)/(Ffi)} \quad (18)$$

If the egg hatches, choose the worst male and female snakes and swap them out.

$$Sm_{worst} = S_{min} + R \times (S_{max} - S_{min}) \quad (19)$$

$$Sf_{worst} = S_{min} + R \times (S_{max} - S_{min}) \quad (20)$$

The worst snake in the male group is $Sm_{worst}$, while the worst solution in the female group is $Sf_{worst}$. The diversity factor operator pm, which also provides the option of raising or decreasing snake placements, can be used to vary the snake positions in the search space in any direction.

## 4. Background

The binary version produced by SO persistent methods is shown in this section. Additionally, it demonstrates how to approach the FS problem using BSO.

### 4.1. Binary Snake Optimizer (BSO)

FS is a discrete binary optimization problem. Thus, the optimizer must initial- ize the binary snakes in the search space Hashim, Khurma, Albashish, Amin, and Hussien (2023); Hussien, Khurma, Alzaqebah, Amin, and Hashim (2023). You must also ensure that the snakes stay binary when they move to other locations during the update process Khurma et al. (2023). The feature is cho- sen in accordance with binary values when the bit value is 1 or when it is 0. The objective of SO is to simultaneously enhance the number of features selected and classification performance. In order to do this, one must find the best snake with the fewest features and the highest performance score.

The sigmoid function, a typical transfer function used to produce probability values for each snake, is shown in Eq. (21), where $S_d(c)$ denotes the ith snake at iteration c in dimension

d. A bit is given a value of one by the assign- ment function Eq. (21) if the probability exceeds a predetermined threshold and a value of zero if it falls below that threshold.

$$X = \frac{1}{1 + e^{-S_i^d(c)}} \quad (21)$$

$$S_i^d(c+1) \begin{cases} 0, & R < X\ X(S_i^d(c)) \\ 1, & R \geq X\ X(S_i^d(c)) \end{cases} \quad (22) \quad (22)$$

The transfer function shown illustrates Eq. (21), which is used to indicate the likelihood of changing the placements of the elements in the FS technique described in this study.

### 4.2. BSO for Feature Selection

The fitness function is shown in Khurma, Aljarah, Sharieh, and Mirjalili (2020), where Err is the classification error rate, SF is the number of selected features, and AF is the total number of features in a dataset. The dataset is trained using the K-NN technique in this study. The value of the parameter k is 5 Khurma, Aljarah, Sharieh, and Mirjalili (2020). The second crucial aspect of the FS problem is the fitness function. Based on the amount of chosen characteristics and classification error rate as in Eq. (23), snakes are evaluated.

The fitness function may be seen in the citation Khurma, Aljarah, Sharieh, and Mirjalili (2020), where Err denotes the classification error rate, SF the number of features that have been chosen, and AF the total number of fea- tures in a dataset. In this work, the dataset is trained using the K-NN method. The parameter k has a value of 5 Khurma, Aljarah, Sharieh, and Mirjalili (2020). The fitness function is the second important component of the FS problem. Snakes are rated based on the number of selected attributes and classification error rate as in Eq. (23).

$$Fitness = \alpha\ x\ Err + \beta |SF|\ /\ |AF| \quad (23)$$

## 5. The proposed approach

The key BSO modification tactics are discussed in this section with the goal to strengthen the phases of exploration and exploitation as well as to establish a better balance state during the search. When SO is employed to enhance the FS process in the wrapper framework, classification performance may be advantageous.

### 5.1. Hybridization with Logarithmic operator

The Moth Flame Optimization (MFO) Abu Khurmaa et al. (2021) algorithm was built using the logarithmic operator, which had the biggest influence on improving search performance and locating a point within the search space that was extremely close to the ideal snake. The logarithmic operator's benefits include achieving a balance between local and global search (exploration and exploitation). Fig 2 shows the logarithmic spiral motion of Moths in nature

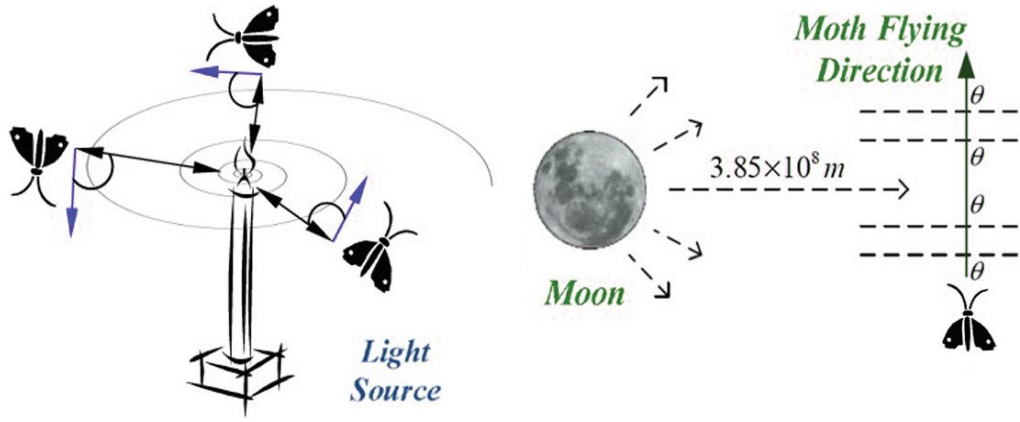

Fig. 2: Logarithmic spiral motion of Moths in nature

The cosine function is used as the foundation for the logarithmic opera- tor, as illustrated in Eq. (24). The distance equation is illustrated in Eq. (25). While fetching the search space, the snake is compelled by the logarithmic function to travel in a spiral pattern. Due to this, the two primary objectives of the optimization method are to have the snakes' positions shift significantly at the start of the research process and gradually throughout the final rounds. Trade-offs between exploration and exploitation are achieved using the adap- tive search method. The second fundamental issue is that it consistently uses the best snake as a search tag. Concerning the top snake, each snake changes its location in the search space. This can be determined by measuring the separation between the present snake and the spiral function's ideal snake.

$$Spiral(CS_i, BS_j) = D_i \times e^{bt} \times cos(2\pi) + BS_j \qquad (24)$$

$CS_i$ is the snake now occupying position ith, while $BS_j$ is the best snake currently occupying position jth. When defining the logarithmic spiral's shape, b and t are constants of value and [ 1, 1] respectively. Option t = 1 indicates how far away the current snake is from the most efficient snake, while t = 1 indicates how far away it is from the least effective snake. To allow for more exploitation in the search area, the t parameter is considered in the range [r, 1], where r drops linearly through rounds from -1 to -2. $D_i$ is defined in Eq. (25).

$$D_i = |Cs_i - Bs_i| \qquad (25)$$

where Di is the distance between the ith present snake and the jth best snake.

Regarding the SO algorithm, the spiral function Eq. (24) and the distance function Eq. (25) will be applied in place of Eq. (10) to make sure that the search space is explored more thoroughly as per the algorithm 1.

---

**Algorithm 1**: Logarithmic Binary Snake Optimizer (LBSO)
---

Input: Dim, UB, LB, N, C, and c
Output: Best Snake
Initialize the binary Snakes randomly
while c ≤ C do
    Assess every solution in categories Nm and Nf

Identify optimal male solution $Sm_{best}$
   Identify optimal female solution $Sf_{best}$
   Identify T based on Eq. (4).
   Define food Q using Eq. (5).
  if (Q < 0.25) then
        Conduct a global search based on Eq. (6) and Eq. (8)
else if (Temperature > 0.6) then
         For i=1 to N
         For j=1 to Dim
         Update r, t
         $D_i = | S(i, j) - L_{food} |$... Eq. (25).
         $S(i, j) = D_i \times e^{bt} \times K(2\pi) + L_{food}$ ...... Eq. (24).
else if (R > 0.6) then
      S in Battle Mode Eq. (11) and Eq. (12)
else
      S in Marriage Mode Eq. (15) and Eq. (16)
      Alternate between the worst male Eq. (19) and female solution Eq. (20)
      Determine transfer function values using Eq. (21)
      Upgrade the solutions using Eq. (22)
      end if
end while
Return the optimal solution

---

### 5.2. Surviving global search schemes

The evolutionary theory, from which the GA algorithm has drawn inspira- tion for numerous swarm-based algorithms, has been included in many earlier works. This combination sped up the convergence process and produced more reliable performance outcomes. The survival concept of the best-fitting solu- tion will be used in this study to force increased diversity among solutions during the exploration phase. Furthermore, rather than picking solutions at random, that might benefit later generations, it will use the principle of natural selection.

When no food is found, snakes in an LSO enter the exploration phase. In other words, if the quantity is less than 0.25, the snakes conduct an extensive global search for a particular target position within the search space. Instead of using SO, they can examine the search space for potential target positions and random positions. This indicates that the target positions may be chosen using the natural selection method. So, the snakes are situated as shown in Eq. (26): One of the Darwinian selection mechanisms, such as proportional selection (or the roulette wheel), is used to choose the position, tournament selection, and linear ranking selection. Therefore, three alternative versions of the LSO have been proposed, namely Tournament LSO (TLSO), Relative LSO (PLSO), and Linear-Rank LSO (LLSO).

$$Sm_i(c + 1) = Sm_t(c) \pm K_2 \times AB_m \times ((S_{max} - S_{min}) \times R + S_{min}) \qquad (26)$$

Where R is a random number in [0, 1], $Sm_{alet}$ is the position of the targeted male snake, and $Sm_i$ is the $i^{th}$ male snake. The male snake's capacity to locate food is represented by $AB_m$ and can be discovered using Eq. (27):

$$ABm = e^{-(F\,mtargeted\,/\,F\,mi)} \quad (27)$$

Where Fitnessmaletargeted is the fitness of Snakemalerand and Fitnessmalei is the fitness of ith snake the in male group and K2 is a constant equals 0.05.

$$Sf_i(c + 1) = Sf_t(c) \pm K_2 \times AB_f \times ((S_{max} - S_{min}) \times R + S_{min}) \quad (28)$$

where $Sf_i$ denotes the position of the chosen female snake, Sft denotes its location, and ABf denotes the capacity of the female snake to locate food. Eq. (9):

$$ABf = e^{-\frac{Fft}{Ffi}} \quad (29)$$

The fitness of the Sftargeted snake is represented by Fftargeted, the fitness of the ith snake in the male group is represented by Ffi, and the constant $K_2$ has a value of 0.05. The suggested SO algorithm's flowchart is shown in Fig 3. The red dashed rectangles represent the new exploration strategy based on the survival of the fittest principle and the new exploitation approach based on the logarithmic cosine operator.

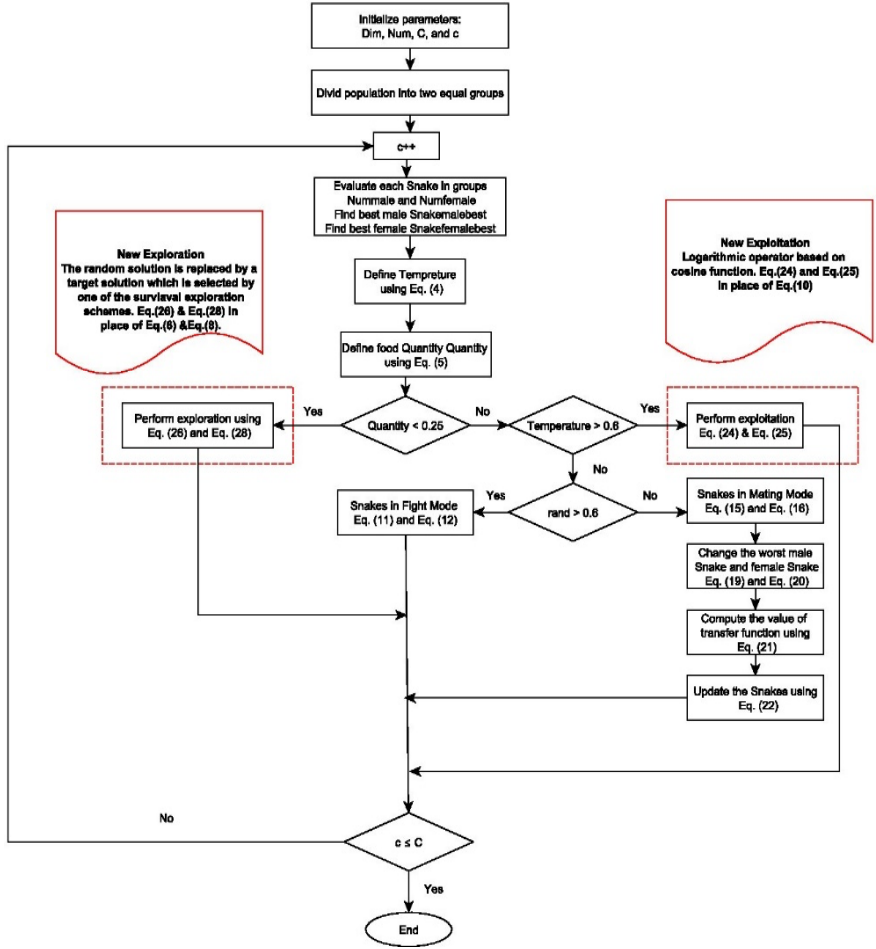

Fig. 3: Logarithmic Survival Exploration Snake Optimizer Flowchart

### 5.2.1. Tournament Logarithmic Snake Optimizer (TLSO)

To carry out the exploration phase of the LSO at once, a tournament selection method is used in this version in place of random selection. Tournament Logarithmic Snake Optimizer (TLSO) is the name of the new version. The main concept is to create a tournament of size o by choosing a group of snakes at random (TS). Typically, o must be less than the sum of N snakes. As demonstrated by Eq. (30), the best snake in TS is then chosen and allocated to $S_o$.

$$S_o = arg\ min_{i\ =\ 1...o} S_{T\ Si} \tag{30}$$

Based on Eq. (31), the selection probability of any snake in TS is determined.

$$P_i\ =\ (1\ /\ N^o)\ ((N-i+1)^o - (N-i)^o) \tag{31}$$

Algorithm 2 shows the steps for implementing tournament survival exploration.

---
**Algorithm 2**: Tournament survival exploration
---
Tindex = random(1, N )
fitness$_1$ = fitness(STindex)
i = 1
    while i <= o do
        Tindex$_2$ = random(1, N )
        fitness$_2$ = fitness(STindex2)
        if (fitness$_2$ < fitness$_1$) then
            Tindex = Tindex$_2$
            fitness$_1$ = fitness$_2$
        end if
        i++
    end while
Target = Tindex
Return Target

---

### 5.2.2. Proportional Logarithmic Snake Optimizer (PLSO)

In this iteration of LSO, the exploration phase of SO is implemented using a proportional selection technique rather than a random one Hassouneh et al. (2021). Proportional Snake Optimizer (PSO) is the name of the new version. The snakes' suitability values served as the foundation for this selecting pro- cess. Each snake's fitness value is related to the overall fitness values of the entire snake when determining the selection probability Pi. The probability of selection for any snake is shown by The Eq. (32):

$$P_i = \frac{F(S_i)}{\sum_{o=1}^{N} F(S_o)} \qquad \forall_i \in (1, 2...N) \tag{32}$$

where

$$F(S_i) = \frac{1}{F(S_i)} \qquad \forall_i \in (1, 2...N) \tag{33}$$

Algorithm 3 shows the steps for implementing proportional survival exploration.

### 5.2.3. Linear-Rank Logarithmic Snake Optimizer (LLSO)

This version of LSO uses a linear order selection scheme instead of random selection in the exploration operator of LSO Al-Betar et al. (2021). The new

---
**Algorithm 3** Proportional survival exploration
---

Calculate the probability of each solution $Prop_j = \frac{fitness(S_j)}{\sum_{k=1}^{N} fitness(S_k)}$

Set Totalprob = 0, Pindex = 0
Generate r ∈ [0, 1]
while Totalprob ≤ r do
   Pindex + +
   Totalprob = Totalprob + Propj
   Target = Pindex
end while
Return Target

---

version is called Linear-Rank Snake Optimizer (LLSO). This selection scheme is a variation of the proportional selection scheme. First, the solutions Ranked as the snake with the best fitness functionality is ranked Firstly; The snake with the second best fitness function comes second, . . ., and the snake with the worst fitness function is ranked last (i.e., N). Next, the selection probability is calculated for each snake relative to its rank as in Eq. (34):

$$F(S_i) = \frac{1}{N} \times (\eta^+ - \eta^-) \times \frac{(i-1)}{(N-1)} \quad (34)$$

where i is the rank of the snake $S_i$. Both $\eta^+$ and $\eta^-$ determine the slope of the linear rank function such that $1 <= \eta^+ <= 2$, and $\eta^- = 2 - \eta^+$. The value of $\eta^+$ refers to the guess value of the snake with the best fitness value $\eta^+ = N \times P_1$ and $\eta^-$ is the guess value of the snake with the worst fitness $\eta^+ = N \times P_N$.
Algorithm 4 shows the steps for implementing linear rank survival exploration.

---
**Algorithm 4** linear rank survival exploration
---

r = random(0, 1), Totalprob = 0
$\eta^+ \in [2, 0]$, $\eta^- = 2\eta^+$
for i=N to 1 do
   $Prop_i = (1/N) \times (\eta^+ - \eta^-) \times (i-1) / (N-1)$
   Totalprob = Totalprob + $Prop_i$
   if (Totalprob ≥ r ) then
      Lindex = i

```
        Break
    end if
    i++
  end for
Target = Lindex
 Return Target
```
---

The Big-O notation was used to indicate the time complexity of the pro- posed BSO (i.e., the worst case). The time complexity analysis of these methods for feature selection tasks is primarily based on the initialization process, dataset dimensions (d), fitness function cost (C), number of optimization algorithm iterations (K), population size (n) (i.e., the number of male + female populations), and number of running experiments (V). The S-shaped transfer function is also utilized to generate binary versions of the BSO. The general computational complexity of the BSO can be expressed using the Big-O case utilizing the notations given above:

$$O\ (BSO) = O\ (init.) + O\ (K\ (pop.\ update)) \\ + O\ (K\ (fitness\ eval.)) + O\ (K\ (selection)) \quad (35)$$

By calculating the Big-O for each phase in Eq. (36), the time complexity for BSO can be represented as the following:

$$O\ (BSO) = O(nf + nm)d + O\ (V\ K(nf + nm)d) \\ + O\ (V\ K(nf + nm)c) + O\ (V\ K(nf + nm)d) \quad (36)$$

It should be noticed that the selection mechanism in of the BLSO requires a single step to select the location of the snake randomly. However, each proposed. version of BLSO (i.e., PLSO, TLSO, and LLSO) require extra time complexity to select a location of the snakes based on the natural selection principle.

For TLSO, the time complexity is further increased according to the tour- nament size T as shown in Algorithm 2 which require:

$$O\ (BSO) = O(nTf + nm)d + O\ (V\ K(nTf + nm)d) \\ + O\ (V\ K(nTf + nm)c) + O\ (V\ K(nTf + nm)d) \quad (37)$$

The PLSO requires [O(n) to select the location of the snake using the code shown in Algorithm 3. Thus the time complexity of PLSO is as follows:

$$O\ (BSO) = (n^2f + n^2m)d + O\ (VK\ (n^2f + n^2m)d) \\ + O\ (VK(n^2f + n^2m)c) + O\ (VK(n^2f + nm)d) \quad (38)$$

Finally, the LLSO requires extra time complexity following the code given in Algorithm 4. Thus the time complexity for LLSO is as follows:

$$O\ (BSO) = O\ (n^2\ logn\ f + n^2\ logn\ x\ m)d + O\ (VK(n^2 logn f + n^2 \times logn \times m)d) + O\ (V\ K(n^2 logn f \\ + n^2 mlogn)c)\ + O\ (V\ K(n^2 logn f + nm)d) \quad (39)$$

It worth mentioning that the time complexity for the objective function for any problem is neglected in the time complexity because it varies concerning the complexity of the optimization problem.

## 6. Experimental Results and Discussions

In this section, we give an overview of the experiments that were conducted and the results obtained to demonstrate the effectiveness of the enhanced binary versions of the snake optimizer that have been suggested when used to pick features in well-known classification tasks. Here, the emphasis is on demonstrating the inherent properties of the suggested algorithms in terms of exploration and exploitation. Along with presenting the specifics of the experimental findings, we also engage in a thorough analysis and fair comparison of the strategies that have been suggested for use in the area of FS for classification enigmas.

### 6.1. Description of the Datasets and evaluation measures

In the experiments provided in this, we have carefully chosen 22 datasets for training and testing the proposed algorithms. Table 1 displays the adopted datasets and points to the number of features and the number of instances available in each dataset. These datasets are highly valued by researchers in the fields of feature selection and classification in that they embrace applications in a variety of fields and vary in their complexity levels from low to medium moving up to high as can be seen from Table 1. This is particularly crucial in analyzing the exploratory and exploitative traits of optimization algorithms utilized in the FS domain. These datasets came from the UCI repository, which is where we got them. We direct interested researchers to Asuncion and Newman (2007) for more details about these datasets. We also want to emphasize that, in our experiments, the enhanced binary versions of the snake optimizer were implemented using the k-NN classifier and a value of 5 was adopted for the Euclidean distance Emary, Zawbaa, and Hassanien (2016).

It is well known that a dataset's inherent characteristics determine its classification or clustering. The middle column of Table 1 lists the low, medium, and high amount of features that each of the 22 datasets possesses. Here, FS methods are crucial for minimizing the number of features used in the classification or clustering processes. The degree of feature duplication determines how the features are chosen. In this circumstance, there is a trade-off since using too few features may impair the accuracy of the classification or clustering while using too many characteristics may render the FS strategy useless. This mission's objective is to reduce the number of features.

The number of instances (also known as samples) available in each of the 22 datasets is indicated in the column that comes before the last column on the right in Table 1. These cases are often split at random into two pools for testing and training the classification system. In the studies carried out for this investigation, there was a 70/20 split between training and testing respectively. Additionally, the $\alpha$ and $\beta$ parameters of the fitness function shown in Eq. (23) were given values of 0.99 and 0.01, respectively. These values have been often cited in numerous published works Khurma et al. (2023).

Table 1: Disease diagnosis benchmark datasets.

| DS_Num | DS_Name | DS_Features | DS_Instances | DS_Classes |
|---|---|---|---|---|
| DS_1 | Diagnostic | 30 | 569 | 2 |
| DS_2 | Original | 9 | 699 | 2 |
| DS_3 | Prognostic | 33 | 194 | 2 |
| DS_4 | Coimbra | 9 | 115 | 2 |
| DS_5 | BreastEW | 30 | 596 | 2 |
| DS_6 | Retinopathy | 19 | 1151 | 2 |
| DS_7 | Dermatology | 34 | 366 | 6 |
| DS_8 | ILPD-Liver | 10 | 583 | 2 |
| DS_9 | Lymphography | 18 | 148 | 4 |
| DS_10 | ParkinsonC | 753 | 755 | 2 |
| DS_11 | SPECT | 22 | 267 | 2 |
| DS_12 | Cleveland | 13 | 297 | 5 |
| DS_13 | HeartEW | 13 | 270 | 2 |
| DS_14 | Hepatitis | 18 | 79 | 2 |
| DS_15 | SAHeart | 9 | 461 | 2 |
| DS_16 | Spectfheart | 43 | 266 | 2 |
| DS_17 | Thyroid0387 | 21 | 7200 | 3 |
| DS_18 | Heart | 13 | 302 | 5 |
| DS_19 | Pima-diabetes | 9 | 768 | 2 |
| DS_20 | Leukemia | 7129 | 72 | 2 |
| DS_21 | Colon | 2000 | 62 | 2 |
| DS_22 | Prostate_GE | 5966 | 102 | 2 |

For more stable results in the carried-out experiments, we have decided to perform thirty independent runs for each experiment. We will shortly show the statistical outcomes of those experiments with 30 independent runs for each. The adopted dimensionalities in our FS algorithms were based upon the number of features related to each dataset listed in Table 1. In our attempt to find the ideal number of features for each dataset associated with the lowest error rate, as revealed by the Euclidean distance value used for the k-NN classifier, we have also presented three binary variations of the classic snake optimizer. We decided to execute each experiment thirty times in total to achieve more reliable results. We'll soon present the statistical findings from those tests, each of which had 30 independent runs. The number of features associated with each dataset indicated in Table 1 served as the basis for the adopted dimensionalities in our FS techniques.

The proposed approaches are evaluated using accuracy, the number of selected features, running time, sensitivity and specificity, convergence curves, and P-value. The following are descriptions of the accuracy, sensitivity, and specificity measures along with their formulas and meaning relative to dis- ease diagnosis. Eq. (40) shows the mathematical formulas of the classification accuracy, Eq. (41) shows the sensitivity , and Eq. (42) shows the specificity.

$$\text{Accuracy} = \frac{TP + TN}{TP + TN + FP + FN} \quad (40)$$

where

– True positive (TP): demonstrates the cases that are truly sick and the classifier anticipates them to be sick.

Table 2: Parameter Settings of the proposed versions of the basic SO.

| Parameter | Value |
|---|---|
| Population size | 50 |
| Number iterations | 100 |

- True negative (TN): demonstrates the cases that are truly in good health and the classifier anticipates them as in good health.
- False positive (FP): demonstrates the cases that are in good health and the classifier anticipates them as sick.
- False negative (FN): demonstrates the cases that are truly sick and the classifier anticipates them as in good health.

$$\text{Sensitivity} = \frac{TP}{TP + FN} \qquad (41)$$

$$\text{Specificity} = \frac{TN}{TN + FP} \qquad (42)$$

### 6.2. Experimental Setup

As noted, the k-NN classifier with a 5 Euclidean distance was used to implement the proposed binary versions of the enhanced SO. The closest training states in the feature space were used as the input for this classifier, and the output corresponds to class membership. The majority of votes are used as the basis for labeling. For each FS problem under consideration, the lower and upper limits were assigned to 0 and 1, respectively. Additionally, the dimensions that were selected for each FS problem were determined by the number of features associated with each dataset that were included in Table 1. The settings of the parameters for the proposed algorithms of SO are shown in Table 2.

For more stable results in the carried-out experiments, we have decided to perform thirty independent runs for each experiment. We will shortly show the statistical outcomes of those experiments with 30 independent runs for each. The adopted dimensionalities in our FS algorithms were based upon the number of features related to each dataset listed in Table 1. The performance of the suggested snake optimization algorithm variants was assessed using independent runs with varied random seeds to demonstrate their stability and robustness in solving FS problems. The research we present here was done on a basic desktop PC with an 11th Gen Intel(R) Core (TM) i5-1135G7@2.40 GHz 2.42 GHz processor and 16.0 GB RAM, and 64-bit operating system, x64-based processor.

The same population size (which corresponds to the number of search agents in this context) and a fixed value of the maximum number of iterations were used for each of the experimented algorithms to unitize the experimental setup for the three proposed binary variants of the standard snake optimization algorithm. The population size adopted was 50, and 100 was selected as the maximum number of iterations. This ensures that the candidate and judge will be compared fairly.

### 6.3. Evaluation of the Proposed FS Algorithms

In this subsection, we compare the three proposed versions of the snake optimizer. These versions go by the names TLSO, PLSO, and LLSO. To harmonize the experimental settings for the proposed competing optimization algorithms of the parent SO, the same population size and the maximum number of iterations were used for each of the competing algorithms.

As stated in Table 2, the adopted population size was 30, and a value of 100 was chosen for the maximum number of iterations. As was already said, this guarantees a fair comparison of the competing algorithms. To get statistically significant results, the experimental runs were carried out 30 times in total, independently. The statistical results were then acquired based on the general capabilities and outcomes from these runs. Each addressed dataset has the same number of features as the dimension of each FS problem that has been solved in this study.

To compare and contrast these algorithms, several performance metrics were used, including the average classification accuracy, average fitness, av- erage sensitivity, average specificity, the average number of features selected, average computational CPU times, and average convergence rates overall runs of the evaluated algorithms. To compare the proposed competing algorithms of SO among themselves, the average (AVG), standard deviation (STD), best, and worst metric measures for each of the performance metrics associated with the offered algorithms used in this study were also used. The algorithms' accuracy level was evaluated using the average measure, and their consistency of performance across different runs was ensured using the standard deviation values. Table 4 displays the comparisons between the different versions of the proposed binary SO algorithm in terms of the mean, worst and best classification accuracy. Table 5 displays mean, worst and best sensitivity results, Table 6 displays mean, worst and best specificity, Table 7 displays mean, worst and best fitness values. Table 9displays mean, worst and best number of selected features. Table 9 displays mean, worst and best CPU execution times. To set them apart from the other results and give them more weight, the best findings are emboldened in all tables.

After a deeper look at Table 4, one can reach the following conclusions. In general, the performance of the three proposed algorithms was excellent to a great extent in terms of the high accuracy figures scored and the low corresponding standard deviations plotted. There are however some variations. Specifically, TLSO superseded its companions exclusively in terms of the obtained accuracy in 14 out of the 22 datasets and scored the highest accuracy in a total of 16 datasets. Similarly, PLSO was the exclusive best in 6 datasets, respectively and has the highest in 8 datasets. Concerning the proposed LLSO, no outcome was reported to be better than others in any of the datasets. A 100% accuracy was not possible in any of the datasets, and accuracy approaching 100% was obtained in several datasets. An accuracy value close to one hundred percent was easily achieved by all the three proposed algorithms in the DS_5, DS_17, DS_20, and DS_22 datasets. Close to that, we can see an accuracy value close to 100% obtained in the DS_1 dataset by TLSO and PLSO, while the accuracy obtained by LLSO was 0.9922%. For the DS_20 datasets, both TLSO and PLSO achieved 99.99% accuracy. TLSO was exclusive in achieving 99.92% and 99.89% accuracy values in DS_5 and DS_17 datasets and PLSO achieved 99.87% accuracy exclusively in the DS_7 datasets. For some datasets like the DS_3, DS_6, DS_11, and DS_15 datasets, it was hard to achieve close-to-80.00% accuracy. This can be attributed to the tough nature of these datasets. Nevertheless, the proposed algorithms achieved comparable, or better, accuracy values than those reported in the literature, as we will see shortly. Regarding the STD values shown in Table 4, either zero or very close to zero values can be observed.

Table 3: The results of BSO without enhancement

| Dataset | Stat. measure | | Accuracy | Sensitivity | Specificity | Time (s) |
|---|---|---|---|---|---|---|
| DS_1 | AVG | | 0.8265 | 0.5287 | 0.9100 | 40.1230 |
| | | STD | 0.0012 | 0.0153 | 0.0145 | 2.1543 |
| DS_2 | AVG | | 0.9413 | 0.9352 | 0.9147 | 59.5122 |
| | | STD | 0.1245 | 0.1289 | 0.0416 | 3.5941 |
| DS_3 | AVG | | 0.7446 | 0.9325 | 0.7129 | 69.2413 |
| | | STD | 0.0597 | 0.3124 | 0.5102 | 5.1489 |
| DS_4 | AVG | | 0.7541 | 0.5847 | 0.7539 | 30.2143 |
| | | STD | 0.9412 | 0.0452 | 0.0014 | 3.2331 |
| DS_5 | AVG | | 0.8224 | 0.8447 | 0.9462 | 22.9482 |
| | | STD | 0.0145 | 0.0015 | 0.0597 | 0.2154 |
| DS_6 | AVG | | 0.7082 | 0.5814 | 0.5846 | 137.2453 |
| | | STD | 0.0124 | 0.0453 | 0.2154 | 22.6421 |
| DS_7 | AVG | | 0.7724 | 0.8854 | 0.6258 | 25.6497 |
| | | STD | 0.2140 | 0.0142 | 0.8512 | 2.6215 |
| DS_8 | AVG | | 0.7951 | 0.6541 | 0.6542 | 45.2143 |
| | | STD | 0.0154 | 0.0149 | 0.1124 | 3.97412 |
| DS_9 | AVG | | 0.9252 | 0.5502 | 0.6558 | 50.2143 |
| | | STD | 0.1543 | 0.0984 | 0.0512 | 44.2153 |
| DS_10 | AVG | | 0.7958 | 0.4157 | 0.7284 | 100.5123 |
| | | STD | 0.0145 | 0.0154 | 0.0125 | 4.21581 |
| DS_11 | AVG | | 0.8015 | 0.4954 | 0.3235 | 80.2145 |
| | | STD | 0.0129 | 0.0111 | 0.0563 | 19.1124 |
| DS_12 | AVG | | 0.5012 | 0.3158 | 0.2214 | 24.2541 |
| | | STD | 0.0124 | 0.0124 | 0.0001 | 3.2154 |
| DS_13 | AVG | | 0.8501 | 0.7913 | 0.8154 | 44.2136 |
| | | STD | 0.0123 | 0.0112 | 0.0145 | 29.1214 |
| DS_14 | AVG | | 0.8224 | 0.7412 | 0.5554 | 30.2147 |
| | | STD | 0.0614 | 0.0254 | 0.0661 | 2.1453 |
| DS_15 | AVG | | 0.7794 | 03954 | 0.5741 | 82.2145 |
| | | STD | 0.0564 | 0.0118 | 0.8741 | 6.14782 |
| DS_16 | AVG | | 0.8145 | 0.5874 | 0.6365 | 75.1547 |
| | | STD | 0.0661 | 0.0217 | 0.0323 | 2.1547 |
| DS_17 | AVG | | 0.9440 | 0.6854 | 0.8695 | 140.2954 |
| | | STD | 0.0100 | 0.0000 | 0.0001 | 5.1243 |
| DS_18 | AVG | | 0.8661 | 0.6144 | 0.7524 | 34.2154 |
| | | STD | 0.0446 | 0.0016 | 0.0549 | 3.2145 |
| DS_19 | AVG | | 0.7557 | 0.5254 | 0.6887 | 79.2355 |
| | | STD | 0.0154 | 0.4125 | 0.0128 | 6.2154 |
| DS_20 | AVG | | 0.8841 | 98254 | 68.951 | 42.5164 |
| | | STD | 0.0214 | 0.1254 | 0.0251 | 0.4514 |
| DS_21 | AVG | | 0.8311 | 0.4154 | 0.7012 | 9.1249 |
| | | STD | 0.0451 | 0.4152 | 0.6513 | 6.2143 |
| DS_22 | AVG | | 0.8219 | 0.3254 | 0.4947 | 190.1250 |
| | | STD | 0.0217 | 0.6325 | 0.4425 | 5.1546 |

Table 4: Comparison of the three binary versions of the snake optimizer in terms of the average classification accuracy measure associated with its STD, best, and worst values.

| Dataset | Mean | | | STD | | | Worst | | | Best | | |
|---|---|---|---|---|---|---|---|---|---|---|---|---|
| | TLSO | PLSO | LLSO | TLSO | PLSO | LLSO | TLSO | PLSO | LLSO | TLSO | PLSO | LLSO |
| DS_1 | 0.9950 | 0.9944 | 0.9922 | 0.0002 | 0.0002 | 0.0002 | 0.9514 | 0.9512 | 0.9487 | 0.9814 | 0.9812 | 0.9810 |
| DS_2 | 0.9691 | 0.9689 | 0.9656 | 0.0000 | 0.0000 | 0.0001 | 0.9401 | 0.9398 | 0.9310 | 0.9745 | 0.9700 | 0.9687 |
| DS_3 | 0.8591 | 0.7711 | 0.7601 | 0.0100 | 0.0010 | 0.0000 | 0.8495 | 0.8410 | 0.8410 | 0.8691 | 0.8779 | 0.8777 |
| DS_4 | 0.9421 | 0.9014 | 0.8847 | 0.0065 | 0.0015 | 0.0022 | 0.9179 | 0.8879 | 0.8759 | 0.9747 | 0.9298 | 0.9211 |
| DS_5 | 0.9992 | 0.9920 | 0.9878 | 0.0003 | 0.0125 | 0.0014 | 0.9587 | 0.9358 | 0.9145 | 0.9854 | 0.9814 | 0.9800 |
| DS_6 | 0.9678 | 0.9254 | 0.9084 | 0.0004 | 0.0024 | 0.0005 | 0.9214 | 0.8210 | 0.8200 | 0.9884 | 0.9349 | 0.9297 |
| DS_7 | 0.9986 | 0.9987 | 0.9874 | 0.0014 | 0.0015 | 0.0020 | 0.9325 | 0.9487 | 0.9312 | 0.9784 | 0.9798 | 0.9784 |
| DS_8 | 0.8904 | 0.8904 | 0.8254 | 0.0142 | 0.0001 | 0.0025 | 0.8145 | 0.8089 | 0.8041 | 0.9757 | 0.9643 | 0.9607 |
| DS_9 | 0.9457 | 0.9457 | 0.9400 | 0.0014 | 0.0147 | 0.0014 | 0.9254 | 0.9378 | 0.9145 | 0.9658 | 0.9698 | 0.9598 |
| DS_10 | 0.8837 | 0.8625 | 0.8600 | 0.0214 | 0.0014 | 0.0051 | 0.8147 | 0.8140 | 0.8045 | 0.9545 | 0.9514 | 0.9400 |
| DS_11 | 0.9271 | 0.9289 | 0.9095 | 0.0054 | 0.0215 | 0.0014 | 0.9045 | 0.9098 | 0.8941 | 0.9494 | 0.9499 | 0.9520 |
| DS_12 | 0.8955 | 0.8340 | 0.8047 | 0.0254 | 0.0147 | 0.0002 | 0.8454 | 0.7940 | 0.7798 | 0.9445 | 0.9310 | 0.9087 |
| DS_13 | 0.9258 | 0.9198 | 0.9175 | 0.0316 | 0.0124 | 0.0214 | 0.9015 | 0.8974 | 0.8910 | 0.9458 | 0.9400 | 0.9348 |
| DS_14 | 0.9874 | 0.9876 | 0.9871 | 0.0101 | 0.0215 | 0.0214 | 0.9741 | 0.9799 | 0.9725 | 0.9978 | 0.9997 | 0.9971 |
| DS_15 | 0.9125 | 0.9159 | 0.8971 | 0.0000 | 0.0000 | 0.0000 | 0.9054 | 0.8851 | 0.8054 | 0.9287 | 0.9398 | 0.9343 |
| DS_16 | 0.9589 | 0.9587 | 0.9471 | 0.0142 | 0.0100 | 0.0100 | 0.8974 | 0.8967 | 0.8964 | 0.9887 | 0.9774 | 0.9766 |
| DS_17 | 0.9989 | 0.9958 | 0.9874 | 0.0215 | 0.0147 | 0.0214 | 0.9687 | 0.9668 | 0.9600 | 0.9945 | 0.9932 | 0.9895 |
| DS_18 | 0.9735 | 0.9601 | 0.9487 | 0.0124 | 0.0002 | 0.0002 | 0.9014 | 0.9010 | 0.8989 | 0.9887 | 0.9868 | 0.9854 |
| DS_19 | 0.9212 | 0.9198 | 0.9101 | 0.0000 | 0.0001 | 0.0125 | 0.8714 | 0.8611 | 0.9099 | 0.9654 | 0.9652 | 0.9249 |
| DS_20 | 0.9999 | 0.9999 | 0.9895 | 0.0000 | 0.0100 | 0.0120 | 0.9715 | 0.9700 | 0.9687 | 0.9999 | 0.9998 | 0.9998 |

| Dataset | | | | | | | | | | | | |
|---------|---|---|---|---|---|---|---|---|---|---|---|---|
| DS_21 | 0.9678 | 0.9679 | 0.9387 | 0.0010 | 0.0001 | 0.0001 | 0.9214 | 0.9210 | 0.9200 | 0.9787 | 0.9774 | 0.9665 |
| DS_22 | 0.9987 | 0.9957 | 0.9897 | 0.0213 | 0.0014 | 0.0141 | 0.9879 | 0.9698 | 0.9666 | 0.9945 | 0.9940 | 0.9938 |

The proposed binary algorithms of SO are compared in Table 4 in terms of average achieved sensitivity values associated with their standard deviations, as well as the best and worst results. The best findings are all included in bold in this table to distinguish them from the other results and to give them greater weight.

We then move on to explain the results of the proposed algorithms of SO in comparison to each other. Tables 5 and 6 provide the average sensitivity and specificity results for the developed competing algorithms of SO as well as their standard deviation, best, and worst values. The performance of the algorithm is improved by obtaining higher levels of sensitivity and specificity. As per the sensitivity results, on can observe that the performance of the proposed TLSO is better than all other proposed versions of SO as it obtained the best sensitivity outcomes in all the 22 datasets. A second look at Table 5 reveals that the proposed TLSO successfully achieved substantial sensitivity results with values larger than 96% in 4 datasets, namely D_1, D_2, D_17,

Table 5: Comparison of the three binary versions of the snake optimizer in terms of the average sensitivity measure associated with its STD, best, and worst values.

| Dataset | Mean | | | STD | | | Best | | | Worst | | |
|---------|------|------|------|------|------|------|------|------|------|------|------|------|
| | TLSO | PLSO | LLSO | TLSO | PLSO | LLSO | TLSO | PLSO | LLSO | TLSO | PLSO | LLSO |
| DS_1 | 0.9686 | 0.9635 | 0.9600 | 0.0012 | 0.0041 | 0.0014 | 0.9351 | 0.9365 | 0.9298 | 0.9684 | 0.9641 | 0.9702 |
| DS_2 | 0.9798 | 0.9598 | 0.9482 | 0.0054 | 0.0078 | 0.0471 | 0.9401 | 0.9397 | 0.9359 | 0.9867 | 0.9750 | 0.9741 |
| DS_3 | 0.8487 | 0.7415 | 0.7332 | 0.0016 | 0.0001 | 0.0021 | 0.8274 | 0.8255 | 0.8227 | 0.8674 | 0.8612 | 0.8597 |
| DS_4 | 0.9500 | 0.8298 | 0.8148 | 0.0000 | 0.0470 | 0.0471 | 0.9201 | 0.9145 | 0.9120 | 0.9741 | 0.9824 | 0.9812 |
| DS_5 | 0.9048 | 0.9478 | 0.8742 | 0.0125 | 0.0147 | 0.0047 | 0.9323 | 0.9332 | 0.9312 | 0.9500 | 0.9510 | 0.9500 |
| DS_6 | 0.6947 | 0.6214 | 0.6182 | 0.0214 | 0.0002 | 0.0781 | 0.6210 | 0.6087 | 0.5858 | 0.7642 | 0.7810 | 0.8011 |
| DS_7 | 0.9304 | 0.9145 | 0.9133 | 0.0019 | 0.0000 | 0.0471 | 0.9215 | 0.9054 | 0.8941 | 0.9431 | 0.9601 | 0.9714 |
| DS_8 | 0.7678 | 0.8500 | 0.7400 | 0.0143 | 0.0147 | 0.0547 | 0.8122 | 0.8146 | 0.8001 | 0.8610 | 0.8697 | 0.8714 |
| DS_9 | 0.8654 | 0.8047 | 0.6927 | 0.0171 | 0.0002 | 0.0178 | 0.8201 | 0.8200 | 0.8196 | 0.9057 | 0.9047 | 0.9178 |
| DS_10 | 0.9147 | 0.7882 | 0.7589 | 0.0148 | 0.0147 | 0.0140 | 0.8512 | 0.8397 | 0.8314 | 0.9735 | 0.9947 | 0.9999 |
| DS_11 | 0.7974 | 0.6710 | 0.6183 | 0.0197 | 0.0001 | 0.0017 | 0.7745 | 0.7394 | 0.7329 | 0.8196 | 0.8578 | 0.8532 |
| DS_12 | 0.6400 | 0.5814 | 0.5490 | 0.0014 | 0.0549 | 0.0112 | 0.6001 | 0.6000 | 0.5912 | 0.6845 | 0.6810 | 0.6959 |
| DS_13 | 0.9335 | 0.8914 | 0.8470 | 0.0011 | 0.0147 | 0.0471 | 0.9287 | 0.9097 | 0.8935 | 0.9473 | 0.9687 | 0.9741 |
| DS_14 | 0.8358 | 0.7814 | 0.7489 | 0.0004 | 0.0012 | 0.0158 | 0.8174 | 0.8155 | 0.8100 | 0.8559 | 0.8542 | 0.8503 |
| DS_15 | 0.7897 | 0.6157 | 0.5914 | 0.0147 | 0.0547 | 0.0014 | 0.7489 | 0.7045 | 0.7194 | 0.8201 | 0.8814 | 0.8749 |
| DS_16 | 0.8184 | 0.8185 | 0.7847 | 0.0001 | 0.0014 | 0.0547 | 0.8000 | 0.8012 | 0.7748 | 0.8210 | 0.8264 | 0.8571 |
| DS_17 | 0.9854 | 0.8797 | 0.8573 | 0.0000 | 0.0401 | 0.0012 | 0.9687 | 0.9578 | 0.9547 | 0.9974 | 0.9999 | 0.9999 |
| DS_18 | 0.8598 | 0.7801 | 0.7700 | 0.0042 | 0.0711 | 0.0141 | 0.8397 | 0.8254 | 0.8198 | 0.8894 | 0.8941 | 0.9019 |
| DS_19 | 0.8086 | 0.7740 | 0.7698 | 0.0004 | 0.0147 | 0.0166 | 0.7748 | 0.7648 | 0.7591 | 0.8179 | 0.8299 | 0.8374 |
| DS_20 | 1.0000 | 1.0000 | 1.0000 | 0.0897 | 0.0019 | 0.0194 | 1.0000 | 0.9999 | 0.9999 | 0.9999 | 1.0000 | 1.0000 |
| DS_21 | 0.8201 | 0.5687 | 0.5621 | 0.0147 | 0.0147 | 0.0087 | 0.7915 | 0.7915 | 0.5612 | 0.8574 | 0.8541 | 0.9945 |
| DS_22 | 0.6281 | 0.5591 | 0.5498 | 0.0048 | 0.0047 | 0.0476 | 0.5948 | 0.5932 | 0.5578 | 0.6516 | 0.6578 | 0.6941 |

Table 6: Comparison of the three binary versions of the snake optimizer in terms of the average specificity measure associated with its STD, best, and worst values.

| Dataset | Mean | | | STD | | | Best | | | Worst | | |
|---------|------|------|------|------|------|------|------|------|------|------|------|------|
| | TLSO | PLSO | LLSO | TLSO | PLSO | LLSO | TLSO | PLSO | LLSO | TLSO | PLSO | LLSO |
| D_1 | 0.9612 | 0.9290 | 0.9190 | 0.0102 | 0.0011 | 0.0016 | 0.9440 | 0.9412 | 0.9398 | 0.9842 | 0.9811 | 0.9975 |
| D_2 | 0.9801 | 0.9587 | 0.9400 | 0.0155 | 0.0100 | 0.0012 | 0.9521 | 0.9512 | 0.9500 | 0.9994 | 0.9999 | 0.9999 |
| D_3 | 0.8690 | 0.8047 | 0.8098 | 0.0006 | 0.0001 | 0.0025 | 0.7814 | 0.7795 | 0.7712 | 0.8279 | 0.8312 | 0.8300 |
| D_4 | 0.9014 | 0.8576 | 0.8497 | 0.0000 | 0.0000 | 0.0001 | 0.8574 | 0.8258 | 0.8258 | 0.9578 | 0.9741 | 0.9845 |
| D_5 | 0.9895 | 0.9671 | 0.9605 | 0.0111 | 0.0001 | 0.0004 | 0.9687 | 0.9559 | 0.9557 | 0.9954 | 0.9998 | 0.9997 |
| D_6 | 0.7201 | 0.6487 | 0.6487 | 0.0126 | 0.0005 | 0.0141 | 0.7048 | 0.7045 | 0.7045 | 0.7451 | 0.7423 | 0.7400 |
| D_7 | 0.7987 | 0.7800 | 0.7748 | 0.0011 | 0.0000 | 0.0012 | 0.7411 | 0.7298 | 0.7191 | 0.8512 | 0.8710 | 0.8834 |
| D_8 | 0.8487 | 0.6812 | 0.8152 | 0.0001 | 0.0001 | 0.0001 | 0.8176 | 0.8073 | 0.7765 | 0.8714 | 0.8874 | 0.9178 |
| D_9 | 0.8578 | 0.9095 | 0.7408 | 0.0175 | 0.0002 | 0.0104 | 0.8610 | 0.8618 | 0.8601 | 0.9412 | 0.9425 | 0.9400 |
| D_10 | 0.9712 | 0.8425 | 0.8091 | 0.0001 | 0.0001 | 0.0001 | 0.9371 | 0.9345 | 0.9311 | 0.9997 | 0.9997 | 0.9998 |
| D_11 | 0.6892 | 0.6254 | 0.5846 | 0.0124 | 0.0021 | 0.0012 | 0.6555 | 0.6222 | 0.6094 | 0.7148 | 0.7445 | 0.7697 |
| D_12 | 0.6987 | 0.6941 | 0.6758 | 0.0004 | 0.0500 | 0.0014 | 0.6538 | 0.6432 | 0.6402 | 0.7398 | 0.7451 | 0.7425 |
| D_13 | 0.9654 | 0.9187 | 0.8751 | 0.0000 | 0.0000 | 0.0000 | 0.9367 | 0.9187 | 0.9020 | 0.9987 | 0.9997 | 0.9998 |
| D_14 | 0.8512 | 0.8579 | 0.7917 | 0.0014 | 0.0004 | 0.0104 | 0.8047 | 0.8199 | 0.8047 | 0.8842 | 0.8975 | 0.8811 |
| D_15 | 0.8039 | 0.8324 | 0.8014 | 0.0122 | 0.0501 | 0.0020 | 0.8011 | 0.8046 | 0.7697 | 0.8621 | 0.8647 | 0.9010 |
| D_16 | 0.8247 | 0.7714 | 0.7587 | 0.0018 | 0.0002 | 0.0002 | 0.8057 | 0.8053 | 0.8001 | 0.8475 | 0.8421 | 0.8410 |
| D_17 | 0.9810 | 0.9647 | 0.9480 | 0.0000 | 0.0000 | 0.0000 | 0.9569 | 0.9425 | 0.9359 | 0.9997 | 0.9875 | 0.9841 |
| D_18 | 0.8985 | 0.8621 | 0.8598 | 0.0000 | 0.0000 | 0.0000 | 0.8558 | 0.8554 | 0.8554 | 0.9487 | 0.9481 | 0.9465 |
| D_19 | 0.8854 | 0.8514 | 0.8500 | 0.0013 | 0.0100 | 0.0001 | 0.8397 | 0.8388 | 0.8369 | 0.9387 | 0.9357 | 0.9300 |

| | | | | | | | | | | | |
|---|---|---|---|---|---|---|---|---|---|---|---|
| D_20 | 1.0000 | 1.0000 | 1.0000 | 0.0006 | 0.0006 | 0.0006 | 0.9999 | 0.9999 | 0.9999 | 1.0000 | 1.0000 | 1.0000 |
| D_21 | 0.8357 | 0.7514 | 0.7510 | 0.0000 | 0.0000 | 0.0000 | 0.8197 | 0.8095 | 0.8088 | 0.8574 | 0.8645 | 0.8641 |
| D_22 | 0.7810 | 0.6899 | 0.6801 | 0.0002 | 0.0002 | 0.0002 | 0.7412 | 0.7400 | 0.7365 | 0.8247 | 0.8212 | 0.8374 |

and D_20. A hundred percent sensitivity was easily achieved by all the three proposed algorithms, TLSO, PLSO, and LLSO, in the D_20 datasets. Close to that, we can see a more than 95.00% sensitivity obtained in the D_2 datasets by TLSO, and PLSO, while the sensitivity obtained by LLSO was around 94.00%. Also, it is clear that TLSO is placed top since it got the greatest specificity findings in all the 22 datasets, as shown in Table 6, while neither the PLSO algorithm nor the LLSO algorithm reported any best specificity score in any dataset.

Table 7: Comparison of the three binary versions of the snake optimizer in terms of the average fitness measure associated with its STD, best, and worst values.

| Dataset | Mean | | | STD | | | Worst | | | Best | | |
|---|---|---|---|---|---|---|---|---|---|---|---|---|
| | TLSO | PLSO | LLSO | TLSO | PLSO | LLSO | TLSO | PLSO | LLSO | TLSO | PLSO | LLSO |
| DS_1 | 0.1300 | 0.1600 | 0.200 | 0.0012 | 0.0012 | 0.0011 | 0.1500 | 0.1800 | 0.2100 | 0.1100 | 0.1400 | 0.1900 |
| DS_2 | 0.0100 | 0.0170 | 0.5000 | 0.0000 | 0.0000 | 0.1100 | 0.0100 | 0.0190 | 0.7000 | 0.0150 | 0.0140 | 0.3000 |
| DS_3 | 0.2400 | 0.2800 | 0.3200 | 0.0100 | 0.0010 | 0.0000 | 0.2670 | 0.2912 | 0.3894 | 0.2210 | 0.2734 | 0.2401 |
| DS_4 | 0.0300 | 0.1000 | 0.1700 | 0.0065 | 0.2015 | 0.2000 | 0.0909 | 0.1000 | 0.1992 | 0.0122 | 0.0999 | 0.1543 |
| DS_5 | 0.1700 | 0.0400 | 0.1900 | 0.0004 | 0.0024 | 0.0004 | 0.2074 | 0.0702 | 0.2029 | 0.1598 | 0.0112 | 0.1892 |
| DS_6 | 0.3600 | 0.3500 | 0.4000 | 0.0004 | 0.0024 | 0.0004 | 0.3874 | 0.3752 | 0.4399 | 0.3498 | 0.0331 | 0.3722 |
| DS_7 | 0.1800 | 0.2000 | 0.2900 | 0.0023 | 0.0021 | 0.0022 | 0.2210 | 0.2541 | 0.3247 | 0.1610 | 0.1729 | 0.2604 |
| DS_8 | 0.2200 | 0.2500 | 0.2700 | 0.1002 | 0.0011 | 0.0006 | 0.2359 | 0.2849 | 0.3009 | 0.2073 | 0.2227 | 0.2473 |
| DS_9 | 0.1900 | 0.2200 | 0.2800 | 0.0015 | 0.0167 | 0.0018 | 0.0487 | 0.0486 | 0.0574 | 0.0543 | 0.0670 | 0.0602 |
| DS_10 | 0.1800 | 0.2000 | 0.2800 | 0.0004 | 0.0004 | 0.0004 | 0.2097 | 0.2000 | 0.3151 | 0.1691 | 0.1998 | 0.2574 |
| DS_11 | 0.2600 | 0.2900 | 0.2400 | 0.0059 | 0.0215 | 0.0211 | 0.2841 | 0.3042 | 0.2978 | 0.2401 | 0.2810 | 0.1922 |
| DS_12 | 0.3900 | 0.3900 | 0.4200 | 0.0394 | 0.0147 | 0.0142 | 0.4301 | 0.4055 | 0.4598 | 0.3524 | 0.3813 | 0.3994 |
| DS_13 | 0.1200 | 0.1400 | 0.1500 | 0.0056 | 0.0144 | 0.0146 | 0.1448 | 0.1510 | 0.1758 | 0.1054 | 0.1332 | 0.1324 |
| DS_14 | 0.1800 | 0.1700 | 0.1800 | 0.0101 | 0.1005 | 0.1541 | 0.1841 | 0.1998 | 0.2045 | 0.1480 | 0.1502 | 0.1657 |
| DS_15 | 0.2600 | 0.3000 | 0.3800 | 0.0000 | 0.0000 | 0.0001 | 0.2869 | 0.3321 | 0.3911 | 0.2473 | 0.2725 | 0.3734 |
| DS_16 | 0.2500 | 0.2500 | 0.2600 | 0.0000 | 0.0100 | 0.0000 | 0.2972 | 0.3113 | 0.2921 | 0.2182 | 0.1971 | 0.2715 |
| DS_17 | 0.2600 | 0.2600 | 0.2900 | 0.0015 | 0.0111 | 0.0156 | 0.2902 | 0.2859 | 0.3300 | 0.2328 | 0.2411 | 0.2520 |
| DS_18 | 0.1800 | 0.2400 | 0.2500 | 0.0118 | 0.0004 | 0.0001 | 0.1920 | 0.2709 | 0.2734 | 0.1728 | 0.2141 | 0.2335 |
| DS_19 | 0.2600 | 0.2000 | 0.2700 | 0.0006 | 0.0006 | 0.0015 | 0.2765 | 0.2274 | 0.2967 | 0.2567 | 0.1900 | 0.2540 |
| DS_20 | 0.0002 | 0.0100 | 0.0100 | 0.0000 | 0.0100 | 0.0125 | 0.0002 | 0.0101 | 0.1110 | 0.1901 | 0.0100 | 0.0100 |
| DS_21 | 0.2800 | 0.2800 | 0.3700 | 0.0001 | 0.0001 | 0.0000 | 0.3298 | 0.3898 | 0.2571 | 0.4126 | 0.1848 | 0.3332 |
| DS_22 | 0.3900 | 0.4900 | 0.5210 | 0.0278 | 0.0143 | 0.0010 | 0.4250 | 0.5290 | 0.5310 | 0.3641 | 0.4642 | 0.5110 |

In 9 datasets, D_1, D_2, D_4, D_5, D_9, D_10, D_13, D_17, and D_20, the proposed TLSO produced extremely sensible specificity results with values greater than 90%. Moreover, the proposed TLSO has achieved specificity ratings of 98.01%, 98.95%, and 98.10% in the D_2, D_5, and D_17 datasets, respectively. So, based on the sensitivity and specificity findings in Tables 5 and 6, one can say that the proposed TLSO algorithm outperforms the other comrade proposed algorithms.

Optimization algorithms aim at minimizing objective functions. This is embodied in Table 7 as the mean fitness values correlated with their STD, best, and worst results. As appear in Table 7, the three proposed algorithms show low averages of fitness values. Astonishingly, TLSO obtained the minimum average fitness values in 16 datasets exclusively, and shared a minimum fitness value of 0.060 with PLSO in the DS_5 datasets. PLSO obtained the minimum average fitness values in 2 datasets exclusively, namely DS_7 and DS_15, and shared a minimum fitness value with PLSO in the DS_5 datasets. Finally, LLSO obtained the minimum average fitness values in the DS_19, DS_20, and DS_22 datasets exclusively.

Although classification accuracy is very important for any classification algorithm, the number of attributes that are used in the classification process is just as important for any FS algorithm. In light of this, Table 8 shows the average number of attributes used in the classification of the 22 datasets for each of the three proposed algorithms of the basic snake optimization algorithm. In this table, one can notice a greater variation when the three algorithms are contrasted. TLSO was the most successful algorithm in reducing the classification attributes, where the average number of selected attributes was the minimum among the three proposed algorithms of SO exclusively in 15 out of 22 datasets and shared the minimum average number of selected attributes with PLSO in the DS_4, DS_19, and DS_20 datasets, as well as with LLSO in the DS_20 datasets. PLSO had the exclusive minimum of attributes in the DS_11, DS_14, and DS_16 datasets, shared the minimum for the DS_4, DS_19, and DS_20 datasets with TLSO, and shared the minimum for the DS_15 and DS_20 datasets with LLSO, which did not exclusively obtain the minimum number of attributes in any of the datasets. From another perspective, the time it takes an algorithm to finish computations is of paramount importance. In light of this, FS algorithms appreciate short execution times. To shed light on this important parameter, we have measured the CPU execution time of the three proposed algorithms of the snake optimization algorithm in seconds and recorded the results in Table 9.

Table 8: Comparison of the three binary versions of the snake optimizer in terms of the average number of selected features measure associated with its STD, best, and worst values.

| Dataset | Mean | | | STD | | | Best | | | Worst | | |
|---|---|---|---|---|---|---|---|---|---|---|---|---|
| | TLSO | PLSO | LLSO | TLSO | PLSO | LLSO | TLSO | PLSO | LLSO | TLSO | PLSO | LLSO |
| DS_1 | 10.000 | 13.510 | 13.987 | 1.9540 | 2.3392 | 1.2475 | 9.0000 | 10.010 | 10.000 | 11.000 | 16.000 | 16.146 |
| DS_2 | 2.0000 | 3.3000 | 3.4981 | 0.4421 | 0.3780 | 0.3157 | 2.0000 | 2.0000 | 3.0211 | 4.0000 | 4.0000 | 3.2222 |
| DS_3 | 14.200 | 14.350 | 15.147 | 2.2509 | 2.6352 | 2.8451 | 11.000 | 11.000 | 14.302 | 17.000 | 17.000 | 18.012 |
| DS_4 | 5.0000 | 5.0000 | 6.0000 | 0.2000 | 0.2914 | 0.2111 | 5.0000 | 5.0000 | 4.0510 | 6.0000 | 6.0000 | 7.5148 |
| DS_5 | 11.771 | 14.000 | 15.000 | 1.5417 | 2.5784 | 1.6715 | 11.000 | 12.000 | 13.599 | 17.000 | 16.000 | 17.250 |
| DS_6 | 9.0100 | 9.2500 | 9.9544 | 1.8710 | 1.1201 | 1.0124 | 7.0000 | 7.0000 | 9.4751 | 17.000 | 17.000 | 18.240 |
| DS_7 | 13.950 | 15.930 | 16.010 | 1.6574 | 1.5740 | 1.5478 | 12.000 | 13.000 | 15.954 | 14.000 | 14.000 | 16.050 |
| DS_8 | 2.9500 | 3.0000 | 3.0000 | 0.4797 | 0.4451 | 0.6684 | 2.0000 | 2.0000 | 3.0000 | 4.0000 | 5.0000 | 4.8215 |
| DS_9 | 8.0000 | 8.3800 | 8.9975 | 1.2000 | 3.0000 | 3.0000 | 6.0000 | 14.000 | 8.0000 | 10.000 | 13.000 | 10.515 |
| DS_10 | 493.52 | 495.44 | 495.55 | 20.000 | 20.500 | 20.000 | 442.00 | 445.00 | 495.00 | 498.00 | 499.00 | 499.15 |
| DS_11 | 11.540 | 11.500 | 11.999 | 1.5784 | 1.4215 | 1.4987 | 9.0000 | 7.0000 | 10.594 | 14.000 | 16.000 | 12.025 |
| DS_12 | 5.1300 | 5.3300 | 6.0012 | 0.8874 | 0.8945 | 0.9854 | 5.0000 | 5.0000 | 6.0000 | 8.0000 | 8.0000 | 8.1453 |
| DS_13 | 6.9800 | 7.0000 | 7.0000 | 0.1472 | 1.4000 | 1.4283 | 6.0000 | 5.0000 | 6.5641 | 9.0000 | 7.0000 | 8.2140 |
| DS_14 | 6.8100 | 5.8600 | 7.2100 | 1.8974 | 1.7451 | 1.7002 | 4.0000 | 2.0000 | 6.0012 | 7.0000 | 8.0000 | 8.0214 |
| DS_15 | 2.8300 | 2.0000 | 2.0000 | 0.5102 | 0.5210 | 0.5789 | 2.0000 | 2.0000 | 2.0000 | 4.0000 | 3.0000 | 4.0000 |
| DS_16 | 21.000 | 20.700 | 25.792 | 3.2587 | 3.0472 | 3.1254 | 17.000 | 19.000 | 19.584 | 24.000 | 23.000 | 23.005 |
| DS_17 | 8.9500 | 9.0000 | 10.032 | 0.8712 | 1.7104 | 1.9871 | 8.0000 | 6.0000 | 7.0000 | 11.000 | 10.000 | 14.154 |
| DS_18 | 4.0000 | 4.0100 | 5.0000 | 0.5166 | 1.2146 | 1.2591 | 4.0000 | 4.0000 | 4.0000 | 5.0000 | 6.0000 | 6.0148 |
| DS_19 | 3.9000 | 3.9000 | 6.0000 | 0.0000 | 3.9000 | 0.0000 | 3.9000 | 3.9000 | 4.0251 | 4.0000 | 4.0000 | 8.1470 |
| DS_20 | 3472.0 | 3472.0 | 3472.0 | 22.088 | 3442.0 | 38.154 | 3442.0 | 3412.0 | 3435.1 | 3500.0 | 3481.0 | 3482.0 |
| DS_21 | 960.20 | 968.90 | 968.99 | 15.800 | 14.478 | 12.5480 | 954.00 | 949.00 | 950.15 | 979.00 | 987.00 | 985.10 |
| DS_22 | 2819.8 | 2821.6 | 2826.0 | 14.0000 | 15.098 | 15.0000 | 2811.0 | 2800.0 | 2800.0 | 3050.0 | 3050.0 | 3054.4 |

Table 9: Comparison of the three binary versions of the snake optimizer in terms of the average running time measure (in seconds) associated with its STD, best, and worst values.

| Dataset | Mean | | | STD | | | Best | | | Worst | | |
|---|---|---|---|---|---|---|---|---|---|---|---|---|
| | TLSO | PLSO | LLSO | TLSO | PLSO | LLSO | TLSO | PLSO | LLSO | TLSO | PLSO | LLSO |
| DS_1 | 19.1000 | 22.0215 | 24.0147 | 0.0001 | 0.0001 | 0.0001 | 18.97112 | 20.1470 | 24.0014 | 19.22888 | 23.896 | 24.028 |
| DS_2 | 20.9741 | 49.5745 | 51.0152 | 0.0006 | 0.0004 | 0.0002 | 18.0015 | 45.5521 | 49.8712 | 23.9467 | 53.5969 | 52.1592 |
| DS_3  | 19.0014 | 57.2140 | 48.5010  | 0.0000 | 0.0010 | 0.0000 | 15.0001  | 55.5781 | 46.5487  | 23.0027  | 58.8499  | 50.4533 |
| DS_4  | 25.4100 | 20.8130 | 25.9731  | 0.0015 | 0.0014 | 0.0002 | 22.2874  | 19.2140 | 23.8715  | 28.5326  | 22.412   | 28.0747 |
| DS_5  | 5.9742  | 4.8954  | 13.2548  | 0.0120 | 0.0270 | 0.0101 | 4.9145   | 3.7412  | 10.8971  | 27.2791  | 26.4842  | 125.1595|
| DS_6  | 24.9102 | 25.1792 | 120.8713 | 0.0014 | 0.0014 | 0.0015 | 22.5413  | 23.8742 | 116.5831 | 27.2791  | 26.4842  | 125.1595|
| DS_7  | 17.9755 | 17.7924 | 18.7900  | 0.0002 | 0.0002 | 0.0003 | 15.2584  | 15.2547 | 16.2510  | 20.6926  | 20.3301  | 21.329  |
| DS_8  | 19.7861 | 18.7430 | 19.9845  | 0.1001 | 0.0009 | 0.0006 | 17.2100  | 17.6581 | 17.0000  | 22.3622  | 19.8279  | 22.969  |
| DS_9  | 16.4113 | 16.8742 | 18.7224  | 0.0062 | 0.0067 | 0.0050 | 15.9874  | 16.5891 | 17.7551  | 16.8352  | 17.1593  | 19.6897 |
| DS_10 | 66.8410 | 74.9711 | 80.6735  | 0.0008 | 0.0006 | 0.0006 | 60.5100  | 72.2571 | 76.2540  | 73.172   | 77.6851  | 85.093  |
| DS_11 | 17.8713 | 19.8735 | 40.9756  | 0.0057 | 0.0021 | 0.0228 | 15.8847  | 19.5410 | 33.1569  | 19.8579  | 20.206   | 48.7943 |
| DS_12 | 10.9743 | 12.8710 | 15.9744  | 0.0320 | 0.0120 | 0.0167 | 10.1478  | 10.8584 | 15.5895  | 11.8008  | 14.8836  | 16.3593 |
| DS_13 | 15.5791 | 13.8210 | 13.1254  | 0.0052 | 0.0139 | 0.0140 | 13.0002  | 13.5104 | 13.1003  | 18.158   | 14.1316  | 13.1505 |
| DS_14 | 17.5451 | 17.7897 | 18.2479  | 0.0104 | 0.1009 | 0.1557 | 15.5522  | 15.7806 | 17.2001  | 19.538   | 19.7988  | 19.2957 |
| DS_15 | 18.9743 | 18.5757 | 18.9640  | 0.0005 | 0.0004 | 0.0004 | 15.5541  | 15.9874 | 17.1487  | 22.3945  | 21.164   | 20.7793 |
| DS_16 | 19.1045 | 19.3257 | 29.4311  | 0.0000 | 0.0000 | 0.0000 | 18.1011  | 17.5478 | 26.4425  | 20.1079  | 21.1036  | 32.4197 |
| DS_17 | 117.9713| 100.97  | 93.8712  | 0.0014 | 0.0012 | 0.0015 | 105.9974 | 95.9712 | 92.5794  | 129.9452 | 105.9748 | 95.163  |
| DS_18 | 15.7914 | 15.9744 | 17.9831  | 0.0117 | 0.0004 | 0.0001 | 13.5402  | 14.8734 | 17.5974  | 19.22888 | 23.896   | 24.028  |
| DS_19 | 20.8361 | 42.9723 | 58.2594  | 0.0008 | 0.0008 | 0.0011 | 20.7845  | 39.9877 | 55.2947  | 23.9467  | 53.5969  | 52.1592 |
| DS_20 | 43.6790 | 38.2561 | 38.2996  | 0.0100 | 0.0010 | 0.0112 | 40.6655  | 35.2289 | 38.2006  | 23.0027  | 58.8499  | 50.4533 |
| DS_21 | 4.9714  | 4.9001  | 5.9840   | 0.0001 | 0.0001 | 0.0001 | 4.2547   | 4.2692  | 5.2259   | 20.6926  | 20.3301  | 21.329  |
| DS_22 | 75.2581 | 75.2913 | 79.8412  | 0.0268 | 0.0131 | 0.0014 | 72.5974  | 74.1102 | 77.2478  | 28.5326  | 22.412   | 28.0747 |

The winner among the three proposed competing algorithms is TLSO, which was able to minimize the execution times for 13 out of 22 datasets. PLSO hit the minimum execution time in the DS_4, DS_5, DS_7, DS_8, DS_15, DS_20, and DS_21 datasets. Finally, LLSO reached the minimum execution time in only two datasets, namely DS_13 and DS_17.

Looking at Tables 4, 5, 6, 7, 8, and 9 again, one can see that TLSO ranked first in terms of classification accuracy, PLSO came in the second position, and LLSO is the third. Similarly, in terms of sensitivity and specificity metrics, TLSO ranked first, followed in order by PLSO and LLSO. As for the fitness values, the order is TLSO, PLSO, and LLSO. In terms of the number of attributes, the order is TLSO, PLSO, and LLSO.

Looking back at the results presented in Tables 4-9, one can see that TLSO excelled at maximizing classification accuracy, sensitivity, and specificity, while minimizing fitness, the number of selected features, and execution times. This can be explained that the logarithmic function enhances the exploitation capability of the SO when it is used to optimize the feature space and enhance the classification results in the medical application. Furthermore, the tournament selection operators have the greatest effect on optimizing the exploration process. Hence, a greater balance between exploitation and exploration is achieved. Therefore, in the next subsection, TLSO is compared to well-known optimization techniques reported in related literature.

### 6.4. Convergence Analysis

Figs 4 to 25 show the convergence curves of the three proposed feature selection methods, TLSO, PLSO, and LLSO, for DS_1 through DS_22, respectively, and show these convergence trends based on the fitness values metric.

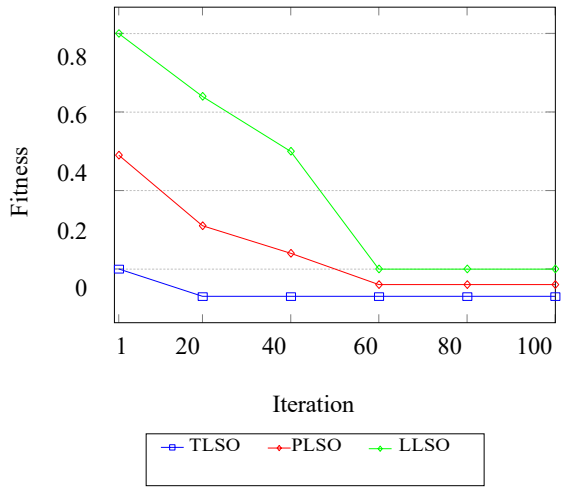

Fig. 4: DS_1

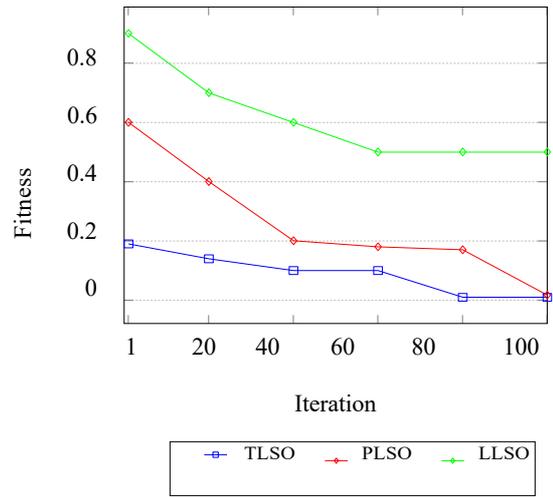

Fig. 5: DS_2

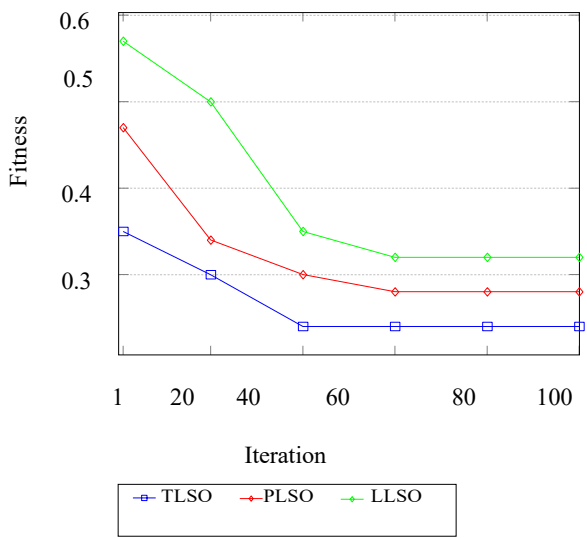

Fig. 6: DS_3

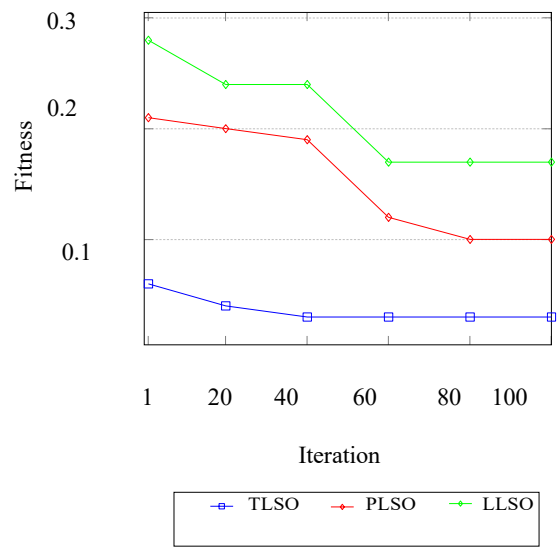

Fig. 7: DS_4

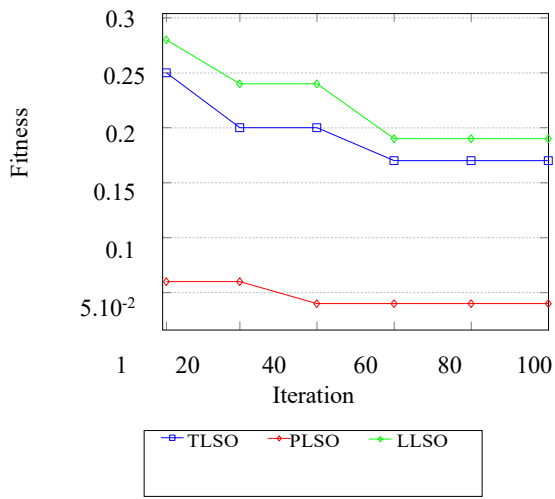

Fig. 8: DS_5

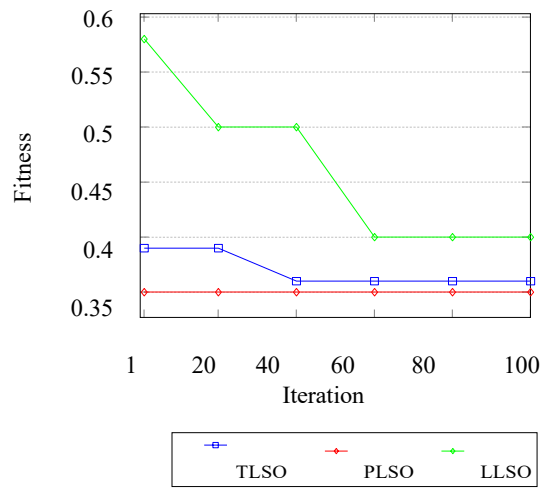

Fig. 9: DS_6

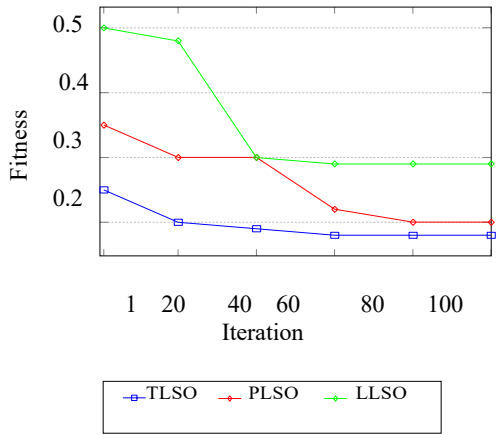

Fig. 10: DS_7

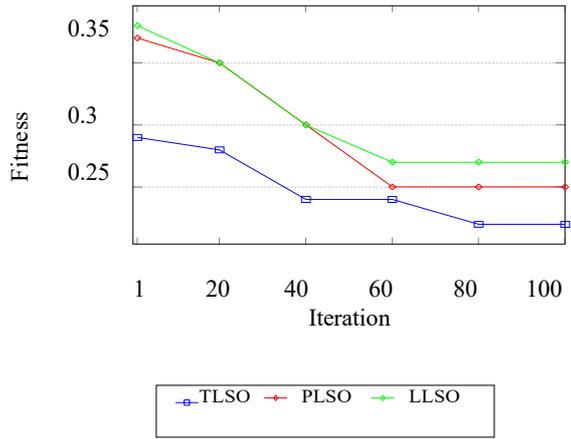

Fig. 11: DS_8

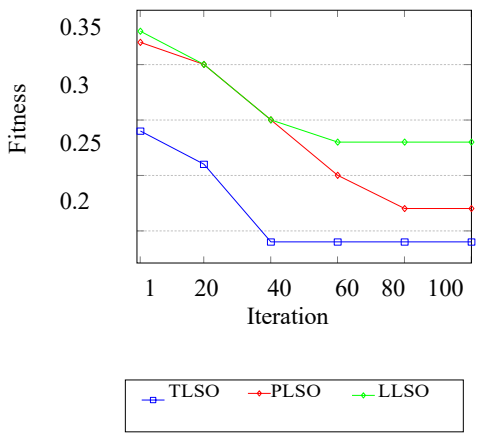

Fig. 12: DS_9

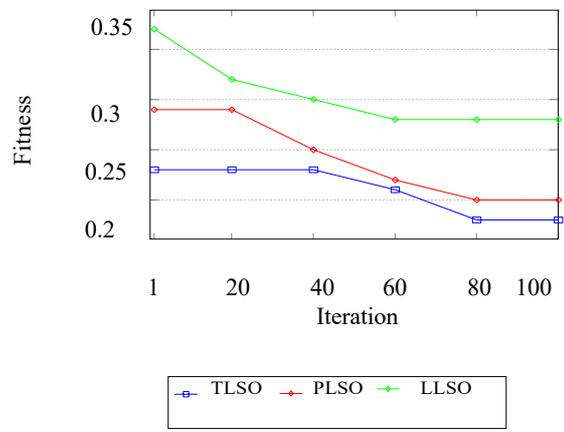

Fig. 13: DS_10

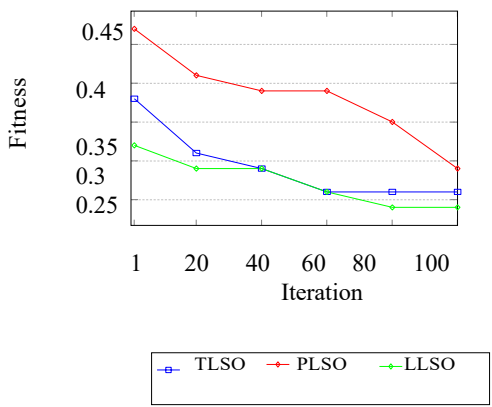

Fig. 14: DS_11

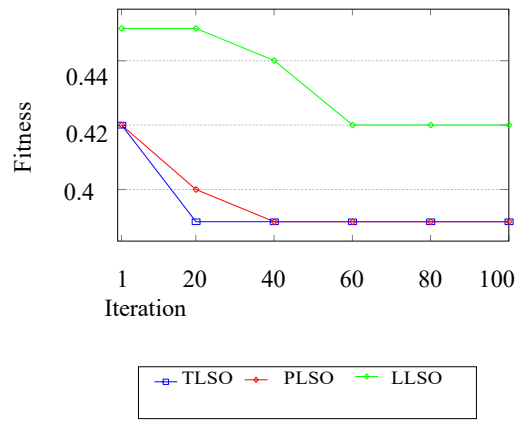

Fig. 15: DS_12

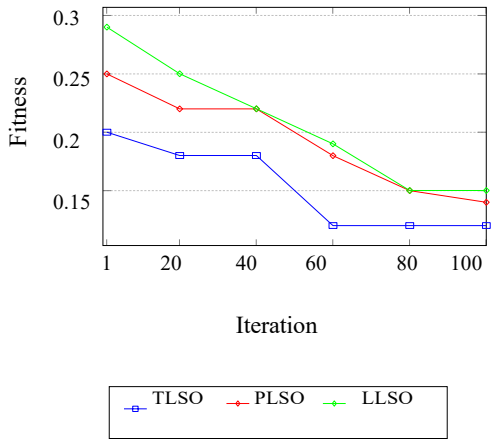

Fig.16: DS_13

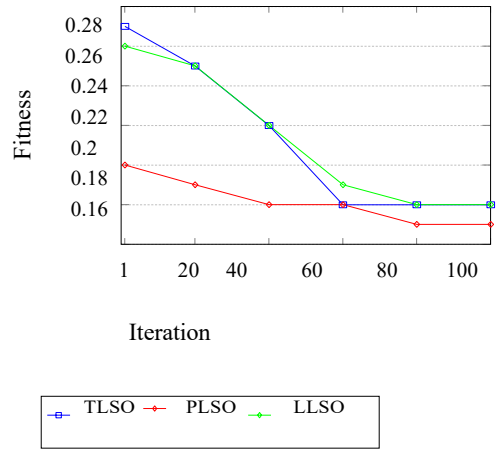

Fig. 17: DS_14

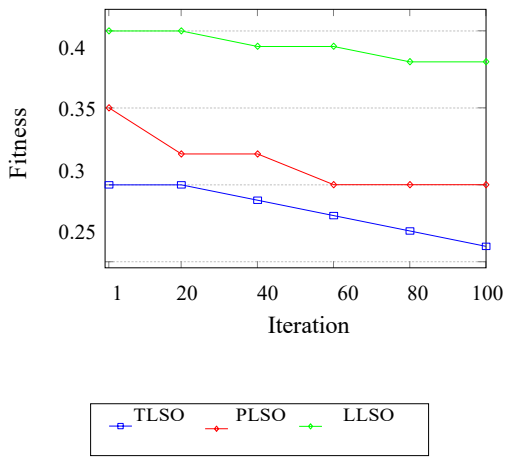

Fig. 18: DS_15

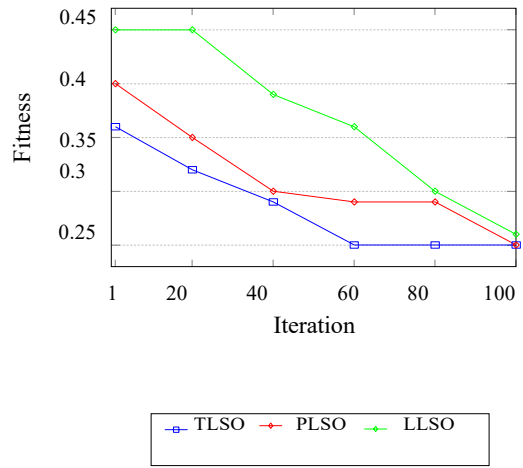

Fig. 19: DS_16

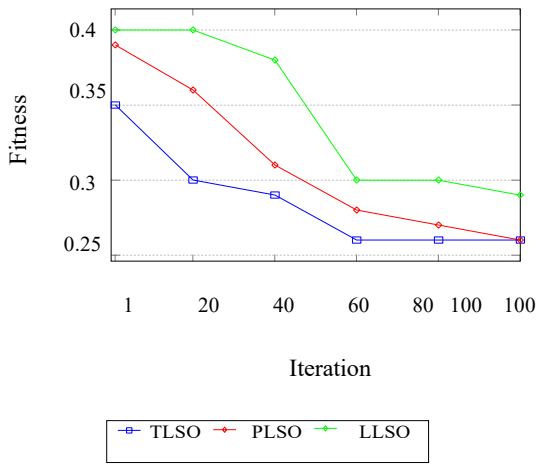

Fig. 20: DS_17

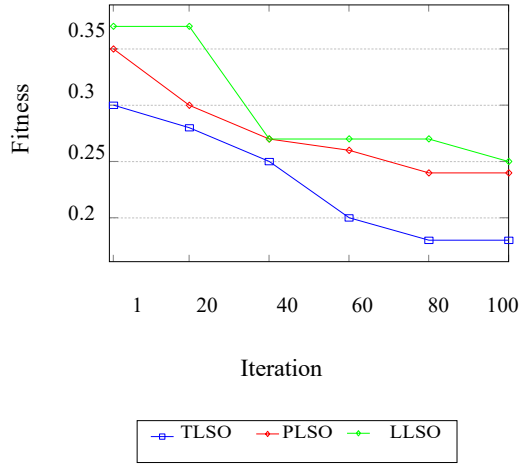

Fig. 21: DS_18

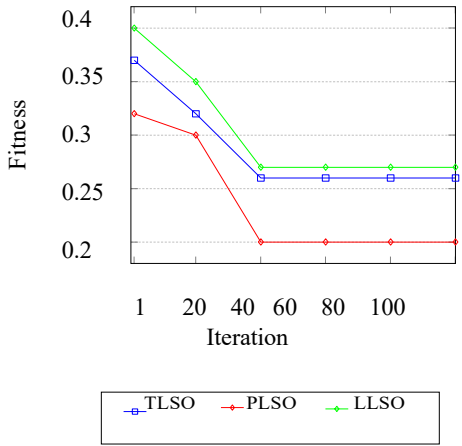

Fig. 22: DS_19

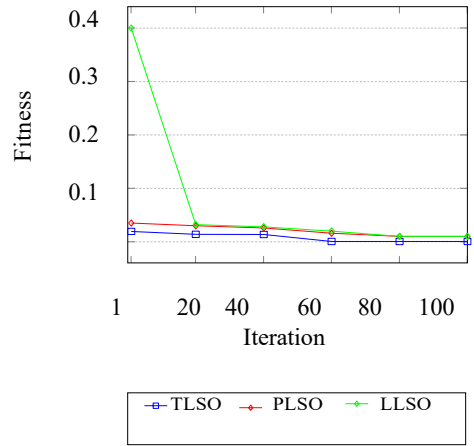

Fig. 23: DS_20

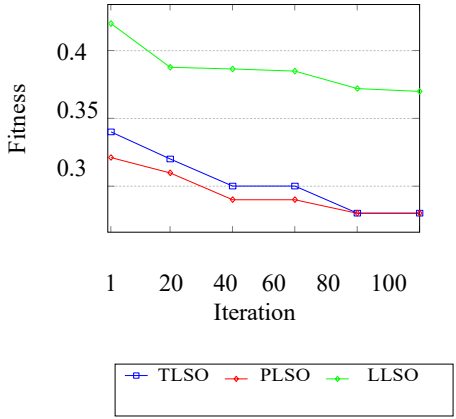

Fig. 24: DS_21

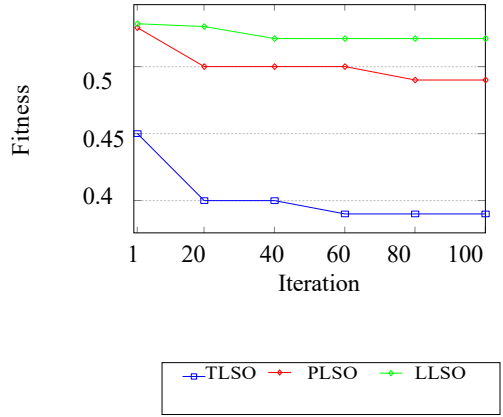

Fig. 25: DS_22

**6.5. Comparison with other Wrapper-based Algorithms**

The performance degree of the developed TLSO algorithm was contrasted with other three binary optimizers including Binary Snake Optimizer with Crossover (BSO-CV) Khurma et al. (2023), Coronavirus Herd Immunity Optimizer (CHIO) with Greedy Crossover (CHIO-GC) Alweshah, Alkhalaileh, Al-Betar, and Bakar (2022), and Binary Moth Flame Optimization with Lévy flight (LBMFO-V3) Alweshah et al. (2022). For fair comparison, these comparative algorithms were carried out using the same parameter settings, including the the maximum number of iterations and the number of search agents employed by the proposed TLSO algorithm.

Performance degrees of the competing feature selection algorithms are compared in respect of the achieved classification accuracy and average number of selected attributes. These comparisons are briefed in respect of the average (AVG) outcomes. The results of the proposed TLSO and other algorithms in regards to the average classification accuracy and number of selected features are given in Table 10. The best results are discriminated in bold to reveal their prominence.

Let us first look at the classification accuracy results presented in Table 10. The superiority of the proposed TLSO is evident from various facets. One can see from Table 10 that TLSO was able to achieve the highest accuracy figures with large margins in 10 out of 22 datasets. The DS_2 - DS_5, DS_7, DS_9, DS_13, DS_14, DS_16, DS_17, DS_20, and

DS_22 datasets were the exception, where, for example, TLSO achieved an accuracy of 96.91% compared to 98.86% best accuracy achieved by BSO-CV with only about 1% deviation. Compared to BSO-CV, LBMFO-V3, and CHIO-G, TLSO was in first place for all 22 datasets. As was mentioned earlier, TLSO was able to achieve 100% accuracy in the DS_20 datasets. The second of the comparisons between the competing algorithms is in terms of the amount of selected features that the algorithm reduces to solve the feature selection problem under study. This comparison is also outlined in Table 10. The division among the four competing algorithms in terms of their average attributes was as follows: TLSO got the least number of attributes in 13 datasets, LBMFO-V3 minimized the number of attributes in 5 datasets, CHIO-GC minimized the number of attributes in 3 datasets, and BSO-CV was the best in one dataset, DS_14. This is shown in Table 10. For the DS_14 dataset, BSO-CV, LBMFO-V3, and CHIO-G reduced the number of attributes to 4. In short, we see here that TLSO was the most efficient feature selection algorithm among the four competitors. In accordance with this, the improvements made to the basic SO are meaningful, and the accuracy of classification results and the amount of feature size have increased and decreased significantly, respectively, compared to the other competitors.

Table 10: Comparison results of TLSO with BSO-CV, LBMFO-V3, and CHIO- G in terms of average classification accuracy and number of selected features.

| Dataset | Avg classification Accuracy | | | | Avg selection size | | | |
|---|---|---|---|---|---|---|---|---|
| | BSO-CV | CHIO-GC | LBMFO-V3 | TLSO | BSO-CV | CHIO-GC | LBMFO-V3 | TLSO |
| DS_1 | 0.9930 | 0.9033 | 0.9100 | 0.9950 | 14.4000 | 13.3700 | 13.9991 | 10.0000 |
| DS_2 | 0.9886 | 0.9710 | 0.9683 | 0.9691 | 3.3000 | 5.1040 | 5.500 | 2.0000 |
| DS_3 | 0.8474 | 0.6716 | 0.9312 | 0.8591 | 15.5000 | 14.6202 | 3.5103 | 14.2000 |
| DS_4 | 0.8182 | 0.8896 | 0.9312 | 0.9421 | 6.1000 | 3.6007 | 3.5103 | 5.0000 |
| DS_5 | 0.9912 | 0.9400 | 0.9398 | 0.9992 | 16.0000 | 13.7303 | 13.9714 | 11.7710 |
| DS_6 | 0.7391 | 0.6436 | 0.5380 | 0.9678 | 11.2000 | 7.2647 | 6.9002 | 9.0100 |
| DS_7 | 1.0000 | 0.8006 | 0.8442 | 0.9986 | 17.3000 | 18.4900 | 18.3541 | 13.9500 |
| DS_8 | 0.7966 | 0.7716 | 0.7143 | 0.8904 | 4.0000 | 4.0000 | 4.0000 | 2.9500 |
| DS_9 | 0.9729 | 0.8343 | 0.8002 | 0.9457 | 9.4000 | 10.0622 | 9.7520 | 8.0000 |
| DS_10 | 0.7873 | 0.8400 | 0.8190 | 0.8837 | 461.4000 | 365.8322 | 369.1070 | 493.5200 |
| DS_11 | 0.8180 | 0.6960 | 0.6576 | 0.9271 | 12.5000 | 9.6050 | 10.7832 | 11.5400 |
| DS_12 | 0.7041 | 0.5966 | 0.5333 | 0.8955 | 6.4000 | 6.8097 | 6.6899 | 5.1300 |
| DS_13 | 0.9370 | 0.9116 | 0.9388 | 0.9258 | 7.3000 | 7.0105 | 6.3100 | 6.9800 |
| DS_14 | 0.9875 | 0.7903 | 0.7500 | 0.9874 | 5.9000 | 8.2011 | 8.3569 | 6.8100 |
| DS_15 | 0.7500 | 0.7036 | 0.6992 | 0.9125 | 3.1000 | 3.1551 | 3.2222 | 2.8300 |
| DS_16 | 0.9084 | 0.7303 | 0.7013 | 0.9589 | 22.7000 | 21.0030 | 20.4598 | 21.0000 |
| DS_17 | 0.9881 | 0.9603 | 0.9776 | 0.9989 | 10.0000 | 8.0116 | 8.4563 | 8.9500 |
| DS_18 | 0.9259 | 0.8126 | 0.7603 | 0.9735 | 5.2000 | 6.1505 | 6.2752 | 4.0000 |
| DS_19 | 0.8182 | 0.7956 | 0.8065 | 0.9212 | 4.0000 | 6.8387 | 6.7612 | 3.9000 |
| DS_20 | 1.0000 | 0.9900 | 1.0000 | 0.9999 | 3502.8000 | 3560.5107 | 3570.7137 | 3472.000 |
| DS_21 | 0.9571 | 0.7176 | 0.6667 | 0.9678 | 970.3000 | 1000.0067 | 991.5551 | 960.2000 |
| DS_22 | 1.0000 | 0.6010 | 0.5056 | 0.9987 | 2969.8000 | 2979.4116 | 2984.7153 | 2819.8000 |

## 6.6. Comparison of TLSO with Filter-based Methods and classical algorithms

In this subsection, we also compare TLSO with another set of algorithms including filter-based algorithms and classical algorithms. Table 11 lists four filter-based algorithms alongside the proposed TLSO. The filter-based algorithms are Chi-square, Relief, CFS, in addition to the IG Khurma, Aljarah, and Sharieh (2021). These algorithms can be categorized according to whether they are supervised or unsupervised, as well as according to whether they use univariate or multivariate classification methods. It should be pointed out that these algorithms fall into the supervised category. Moreover, IG is among the univariate methods, whereas CFS represents a multivariate method.

Looking at Table 11, we can find that the proposed TLSO succeeded in achieving the highest classification accuracy in 19 datasets, whereas Chi-square had the highest accuracy in the DS_11, DS_13, and DS_17 datasets. From these results, one can conclude that the proposed TLSO algorithm is a better classifier than supervised, univariate, and multivariate classification approaches. Table 11 also reveals that filter-based techniques are very weak classifiers in some datasets, with classification accuracy as low as 3.98%. Table 12 shows the outperformance of the TLSO when it is compared with other algorithms namely Genetic algorithm (GA), Particle Swarm Algorithm (PSO), and Differential Evolution (DE). From Table 12, the TLSO achieved the highest results with accuracy of 0.9508. It outperformed other algorithms across 68% of the datasets.

This insures the superiority of the proposed TLSO algorithm over a range of well-defined algorithms popular in the literature. In their entirety, the above results and commentaries demonstrated that the proposed models incorporated into the TLSO were profitable. For TLSO as a FS method, these proposed models combined with TLSO strengthened the global search for this method and improved the solutions slightly. Further, they promoted the local search process and improved the convergence process as well, reaching the optimum solution. Figs. 4 to 25 show the convergence behavior of the three optimization feature selection techniques over a period of 100 iterations stated on the x-axis against the fitness value for the y-axis. The characteristic curves of these figures should be seen in combination with Table 7 in order to maximize their utility. The feature selection method that shows swift convergence in these convergence patterns is recommended to best feature selection algorithm. In other words, we choose the feature selection method that, in the fewest number of iterations, settles on a very low fitness value. The behavior of the three proposed algorithms may be seen to exhibit significant variety when examining these sketched figures. This fluctuation is due to the algorithm's exploitation and exploration behavior, which differs for the same proposed method based on the dataset's characteristics from one dataset to the next. Generally, one can see that the convergence curves reach their equilibrium around or about 80 iterations. In general, TLSO performed encouragingly well in the majority of the datasets. In particular, the DS_1, DS_2, DS_3, DS_4, DS_7, DS_8, DS_9, DS_10, DS_13, DS_15, DS_16, DS_17, DS_18, DS_20, and DS_22 datasets demonstrate the superiority of TLSO. In the DS_5, DS_6, DS_14, and DS_19 datasets, PLSO performed almost as well as or better than its rival algorithms, and LLSO demonstrated strong convergence behavior. Nearly all TLSO and PLSO algorithms for DS_15 and DS_21 reacted similarly, requiring 40 and 80 iterations, respectively, to maintain their minimum values.

The accuracy rate convergence curve of TLSO compared to the other comparative filter-based methods is presented in Fig. 26.

Table 11: Average accuracy of TLSO versus filter-based methods

| Dataset | TLSO | Chi-square | Relief | CFS | IG |
|---|---|---|---|---|---|
| DS_1 | 0.9950 | 0.5714 | 0.9585 | 0.9533 | 0.9349 |
| DS_2 | 0.9691 | 0.9091 | 0.6426 | 0.6860 | 0.6759 |
| DS_3 | 0.8591 | 0.5910 | 0.7727 | 0.7576 | 0.7577 |
| DS_4 | 0.9421 | 0.3846 | 0.6672 | 0.5763 | 0.5578 |
| DS_5 | 0.9992 | 0.9365 | 0.8160 | 0.8029 | 0.8128 |
| DS_6 | 0.9678 | 0.6349 | 0.5036 | 0.4783 | 0.5393 |
| DS_7 | 0.9986 | 0.7250 | 0.7248 | 0.4732 | 0.4021 |
| DS_8 | 0.8904 | 0.7106 | 0.5119 | 0.5223 | 0.5264 |
| DS_9 | 0.9457 | 0.8824 | 0.5886 | 0.5533 | 0.5204 |
| DS_10 | 0.8837 | 0.6593 | 0.6590 | 0.6487 | 0.6376 |
| DS_11 | 0.9271 | 0.9667 | 0.5651 | 0.5508 | 0.5460 |
| DS_12 | 0.8955 | 0.3940 | 0.1181 | 0.0398 | 0.0826 |
| DS_13 | 0.9258 | 0.9334 | 0.6153 | 0.5757 | 0.6202 |
| DS_14 | 0.9874 | 0.7778 | 0.5538 | 0.5857 | 0.6417 |
| DS_15 | 0.9125 | 0.6471 | 0.5024 | 0.5115 | 0.5227 |
| DS_16 | 0.9589 | 0.7000 | 0.6079 | 0.6279 | 0.5551 |
| DS_17 | 0.9989 | 1.0000 | 0.6379 | 0.6955 | 0.9773 |
| DS_18 | 0.9735 | 0.5333 | 0.6317 | 0.5575 | 0.6114 |
| DS_19 | 0.9212 | 0.6905 | 0.5147 | 0.5426 | 0.5264 |
| DS_20 | 0.9999 | 0.7120 | 0.6883 | 0.6759 | 0.6410 |
| DS_21 | 0.9678 | 0.5850 | 0.5641 | 0.5116 | 0.5097 |
| DS_22 | 0.9987 | 0.5042 | 0.5033 | 0.4786 | 0.4421 |
| Average | 0.9508 | 0.7046 | 0.6133 | 0.5887 | 0.5981 |

Table 12: Average accuracy of TLSO versus classical metaheuristic algorithms

| Dataset | TLSO | GA | PSO | DE |
|---|---|---|---|---|
| DS_1 | 0.9950 | 0.9422 | 0.9501 | 0.9181 |
| DS_2 | 0.9691 | 0.9682 | 0.9621 | 0.9542 |
| DS_3 | 0.8591 | 0.8651 | 0.8944 | 0.8100 |
| DS_4 | 0.9421 | 0.9322 | 0.9335 | 0.9110 |
| DS_5 | 0.9992 | 0.9844 | 0.9920 | 0.9736 |
| DS_6 | 0.9678 | 0.9563 | 0.9422 | 0.9256 |
| DS_7 | 0.9986 | 0.9723 | 0.9655 | 0.9436 |
| DS_8 | 0.8904 | 0.9010 | 0.8892 | 0.8846 |
| DS_9 | 0.9457 | 0.9450 | 0.9466 | 0.9424 |
| DS_10 | 0.8837 | 0.8954 | 0.8987 | 0.8874 |
| DS_11 | 0.9271 | 0.9155 | 0.9285 | 0.9045 |
| DS_12 | 0.8955 | 0.8978 | 0.8921 | 0.8890 |
| DS_13 | 0.9258 | 0.9166 | 0.9287 | 0.9030 |
| DS_14 | 0.9874 | 0.9844 | 0.9832 | 0.9811 |
| DS_15 | 0.9125 | 0.9012 | 0.9100 | 0.8977 |
| DS_16 | 0.9589 | 0.9423 | 0.9413 | 0.9399 |
| DS_17 | 0.9989 | 0.9839 | 0.9787 | 0.9236 |
| DS_18 | 0.9735 | 0.9744 | 0.9788 | 0.9655 |
| DS_19 | 0.9212 | 0.9210 | 0.9199 | 0.9094 |
| DS_20 | 0.9999 | 0.9956 | 0.9969 | 0.9844 |
| DS_21 | 0.9678 | 0.9544 | 0.9589 | 0.9412 |
| DS_22 | 0.9987 | 0.9945 | 0.9923 | 0.9921 |
| Average | 0.9508 | 0.9429 | 0.9470 | 0.9270 |

Looking at Fig. 26, we find that TLSO significantly outperformed its accompanying feature selection methods, where it achieved a very high accuracy rate compared to all other competitors. The curve implementing the accuracy rate of TLSO compared to other methods started out large and the curve continued to decrease for the other.

T-test was used for comparing the efficiency of the proposed TLSO, PLSO, and LLSO approaches as shown in Table 13.

From the statistical analysis in Table 13 (Based on a significance threshold of 0.05), the obtained p-values indicate that the final outcomes were statistically significant. The efficiency of the TLSO is slightly higher than the rest proposed approaches.

The average classification accuracy rate of the proposed TLSO compared to other rival wrapper-based methods is shown in Fig. 27.

In Fig. 27, one can obviously notice that the proposed TLSO outperformed in average classification accuracy rates all other rival feature selection methods, namely BSO-CV, CHIO-GC, LBMFO-V3. This indicates its superiority in balancing exploration and exploitation features while searching the all promising regions of the search space for the global optimum solution.

The average amount of selected feature size of the proposed TLSO com- pared to other competing wrapper-based feature selection algorithms methods is presented in Fig. 28.

Looking back at the results showing the average amount of selected features displayed in Table 10 and Fig. 28, we can evidently perceive that TLSO outclassed all the other companion feature selection algorithms in this criterion. More specifically, the proposed TLSO algorithm was superior in reducing the number of features than all others. Therefore, one can claim that TLSO has more promising global and local search processes than other well-known optimization feature selection methods reported in the relevant literature.

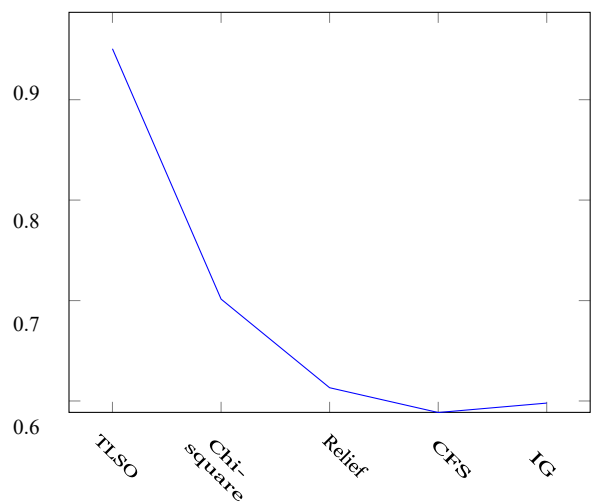

Fig. 26: Accuracy rate of TLSO and other methods.

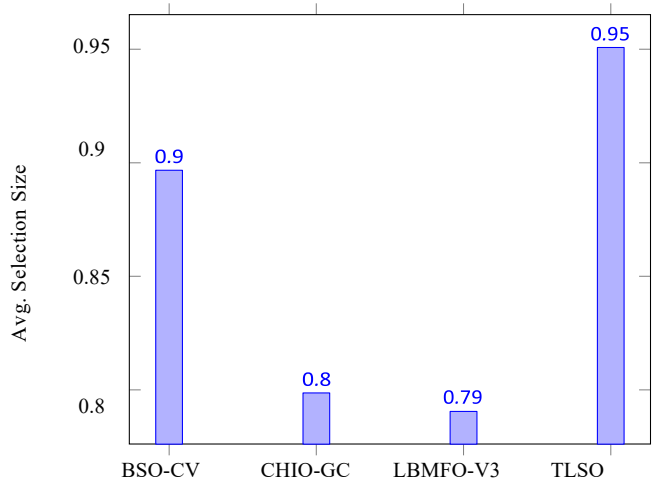

Fig. 27: Average accuracy of BSO-CV, CHIO-GC, LBMFO-V3, and TLSO.

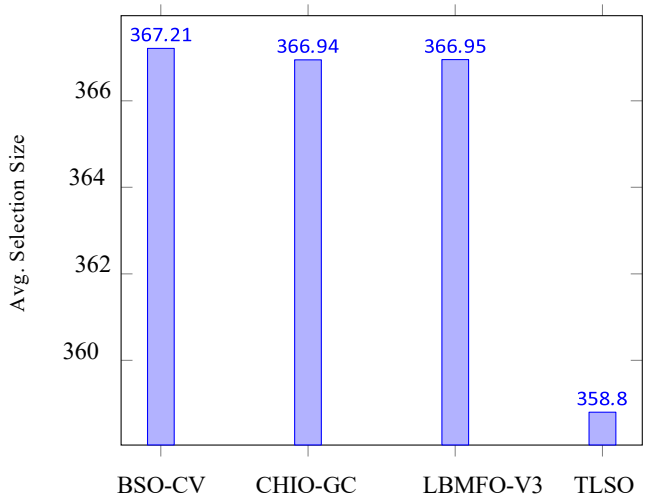

Fig. 28: Average selection size of BSO-CV, CHIO-GC, LBMFO-V3, and TLSO.

Table 13: Ttest for TLSO, PLSO, LLSO.

| Dataset | Method | Avg. | STD | P-Value |
|---|---|---|---|---|
|  | TLSO | 0.9045 | 0.0214 | 0.01 |
| DS_1 | PLSO | 0.9032 | 0.0659 | 0.02 |
|  | LLSO | 0.8821 | 0.0124 | 0.00 |
|  | TLSO | 0.9787 | 0.2541 | 0.00 |
| DS_2 | PLSO | 0.9544 | 0.1599 | 0.00 |
|  | LLSO | 0.9511 | 0.6588 | 0.00 |
|  | TLSO | 0.8745 | 0.3640 | 0.00 |
| DS_3 | PLSO | 0.7869 | 0.1544 | 0.01 |
|  | LLSO | 0.7486 | 0.1487 | 0.02 |
|  | TLSO | 0.8970 | 0.6987 | 0.00 |
| DS_4 | PLSO | 0.7842 | 0.1145 | 0.01 |
|  | LLSO | 0.7798 | 0.1475 | 0.01 |
|  | TLSO | 0.9547 | 0.2547 | 0.03 |
| DS_5 | PLSO | 0.9214 | 0.4573 | 0.01 |
|  | LLSO | 0.9053 | 0.9874 | 0.04 |
|  | TLSO | 0.6657 | 0.5871 | 0.03 |
| DS_6 | PLSO | 0.6582 | 0.7548 | 0.02 |
|  | LLSO | 0.6469 | 0.3154 | 0.01 |
|  | TLSO | 0.8550 | 0.1125 | 0.03 |
| DS_7 | PLSO | 0.8253 | 0.5688 | 0.04 |
|  | LLSO | 0.8247 | 0.5247 | 0.00 |
|  | TLSO | 0.8021 | 0.9874 | 0.01 |

| Dataset | Method | | | |
|---|---|---|---|---|
| DS_8 | PLSO | 0.7914 | 0.1245 | 0.02 |
| | LLSO | 0.7843 | 0.1167 | 0.01 |
| | TLSO | 0.8805 | 0.1547 | 0.00 |
| DS_9 | PLSO | 0.8824 | 0.1987 | 0.01 |
| | LLSO | 0.8741 | 0.4789 | 0.00 |
| | TLSO | 0.8623 | 0.8791 | 0.00 |
| DS_10 | PLSO | 0.8512 | 0.4152 | 0.01 |
| | LLSO | 0.8448 | 0.1197 | 0.00 |
| | TLSO | 0.7787 | 0.1697 | 0.02 |
| DS_11 | PLSO | 0.7512 | 0.4871 | 0.00 |
| | LLSO | 0.7090 | 0.8713 | 0.00 |
| | TLSO | 0.5983 | 0.2297 | 0.01 |
| DS_12 | PLSO | 0.5881 | 0.4810 | 0.02 |
| | LLSO | 0.5812 | 0.8941 | 0.00 |
| | TLSO | 0.9248 | 0.1492 | 0.02 |
| DS_13 | PLSO | 0.9154 | 0.2145 | 0.01 |
| | LLSO | 0.9102 | 0.6124 | 0.01 |
| | TLSO | 0.8120 | 0.5512 | 0.00 |
| DS_14 | PLSO | 0.8111 | 0.4230 | 0.02 |
| | LLSO | 0.8098 | 0.6877 | 0.01 |
| | TLSO | 0.7511 | 0.5014 | 0.00 |
| DS_15 | PLSO | 0.7451 | 0.4180 | 0.01 |
| | LLSO | 0.7423 | 0.1470 | 0.00 |
| | TLSO | 0.7712 | 0.1029 | 0.01 |
| DS_16 | PLSO | 0.7644 | 0.2410 | 0.01 |
| | LLSO | 0.7521 | 0.6512 | 0.00 |
| | TLSO | 0.9784 | 0.1897 | 0.00 |
| DS_17 | PLSO | 0.9684 | 0.2510 | 0.00 |
| | LLSO | 0.9601 | 0.2154 | 0.02 |
| | TLSO | 0.8241 | 0.4321 | 0.01 |
| DS_18 | PLSO | 0.8200 | 0.2141 | 0.02 |
| | LLSO | 0.8198 | 0.2246 | 0.00 |
| | TLSO | 0.8100 | 0.5001 | 0.01 |
| DS_19 | PLSO | 0.8094 | 0.4123 | 0.01 |
| | LLSO | 0.7914 | 0.2411 | 0.00 |
| | TLSO | 0.8461 | 0.8706 | 0.02 |
| DS_20 | PLSO | 0.8423 | 0.7710 | 0.02 |
| | LLSO | 0.8215 | 0.7125 | 0.01 |
| | TLSO | 0.9712 | 0.8712 | 0.00 |
| DS_21 | PLSO | 0.9424 | 0.1230 | 0.00 |
| | LLSO | 0.9345 | 0.1241 | 0.00 |
| | TLSO | 0.6514 | 0.1425 | 0.01 |
| DS_22 | PLSO | 0.6441 | 0.2143 | 0.00 |
| | LLSO | 0.6302 | 0.3244 | 0.00 |

## 7. Conclusion and future research

The outstanding qualities of the SO method, as well as the optimization results obtained by SO in the literature, prompted us to propose it for conducting optimization operations within the wrapper FS framework. Furthermore, some operators in the medical diagnosis industry are required to cope with the SO's inadequacies and increase its classification capacity, such as discriminating between malevolent and benign tumors with high accuracy. Several types of adjustment strategies have been implemented into the SO framework in this study to address these numerous inadequacies. To improve SO exploitation, the logarithmic cosine function from the MFO algorithm is used as the dependent operators. Additionally, to improve SO exploration, proportional and linear schemes and evolutionary selection operators like tournament selection are used. Rather than selecting a solution at random from the swarm and looking for it everywhere. In order to select a course of action and implement the survival of the fittest concept, selection schemes are used. The FS procedure is then enhanced using the improved SO to produce more reliable medical diagnostic outcomes. TLSO, PLSO, and

LLSO are the three contrast models that were developed. The proposed method is evaluated using 22 medical datasets, where the results indicate the effectiveness of the TLSO method in terms of the evaluation measures. The main limitations of this study that we will work on in the future is the scalability which was noticeable when we applied the proposed approach on large dimensionality dataset. This took a large processing time. Hence, in future we can take this limitation into consideration to reduce the processing time on large datasets by applying some mechanisms such as parallelism. Furthermore, for future this topic could be expanded by the researchers using different selection strategies, like exhaustive sampling and exponential ordering. Investigation of TLSO, PLSO, and LLSO techniques for solving complex real-world problems such as image segmentation, cloud resource allocation, and engineering problems could be further study.

### Acknowledgments

This work was supported by the Ministerio Español de Ciencia e Innovación under project number PID2020-115570GB-C22 MCIN/AEI/10.13039/501100011033 and by the Cátedra de Empresa Tecnología para las Personas (UGR-Fujitsu).